\documentclass[%
 reprint,
nofootinbib,
amsmath,amssymb,
aps,twoside
]{revtex4-2}

\usepackage{graphicx}
\usepackage{dcolumn}
\usepackage{bm}
\usepackage{hyperref}
\usepackage{color}
\usepackage[dvipsnames]{xcolor}

\newcommand\authormark[1]{\textsuperscript{#1}}
\newcommand{\std}{\mathrm{SD}}

\newcommand{\vari}{\mathrm{V}}
\newcommand{\lilspace}{\,}

\makeatletter
\renewcommand\@make@capt@title[2]{%
 \@ifx@empty\float@link{\@firstofone}{\expandafter\href\expandafter{\float@link}}%
  {#1}\@caption@fignum@sep#2\quad}%
\makeatother

\makeatletter 
\renewcommand{\fnum@figure}{FIG.~\thefigure}
\makeatother

\begin{document}

\preprint{APS/123-QED}

\title{Quantitative Profilometric Measurement of Magnetostriction in Thin-Films}

\author{Hamish Greenall\authormark{1,2}, Benjamin J. Carey\authormark{1,2}, Douglas Bulla\authormark{4}, James S. Bennett\authormark{3}, Glen I. Harris\authormark{1,2}, Fernando Gotardo\authormark{1,2}, Scott Foster\authormark{4}, and Warwick P. Bowen\authormark{1,2} \email{warwick.bowen@uq.edu.au}}
\address{\authormark{1}School of Mathematics and Physics, The University of Queensland, St Lucia, Queensland 4067, Australia.\\
\authormark{2}ARC Centre of Excellence for Engineered Quantum Systems, St Lucia, Queensland 4067, Australia.\\
\authormark{3}Centre for Quantum Dynamics, Griffith University, Nathan, Queensland 4072, Australia.\\
\authormark{4}Australian Government Department of Defence Science and Technology, Edinburgh,\\ South Australia 5111, Australia.\\
}

\date{\today}

\begin{abstract}
A DC non-contact method for measuring the magnetostrictive strain in thin-films is demonstrated, achieving a state-of-the-art sensitivity of $\mathrm{0.1~ppm}$.  In this method, an optical profilometer is used to measure the curvature induced in a magnetostrictively coated coverslip under a DC field through phase-sensitive interferometry. From this the magnetostrictive stress and strain are calculated using Stoney's formula. This addresses limitations of conventional techniques that measure magnetostriction based on the deflection of a cantilever under an AC field, which require complex dedicated set-ups and are sensitive to vibrational noise. Further, it reveals information about the anisotropy of the film and allows for the possibility of measuring multiple samples simultaneously. The theoretical sensitivity limits are derived, predicting a shot-noise-limit of $0.01~\mathrm{ppm}$. The method is implemented to measure the magnetostrictive hysteresis and piezomagnetic coupling of thin-film galfenol. Degradation in film performance is observed above a thickness of 206~nm, alongside a change in coercivity. This prompts investigation into the growth and optimization of galfenol films for use in devices. 

\end{abstract}
\maketitle
\section{Introduction}

In recent years magnetostrictive materials have been the focus of significant research due to their potential applications in fields ranging from smart materials \cite{wang2023development, zanjanchi2023bifurcation, lu2023cylindrical} to sensing \cite{gotardo2023waveguide,li2012surface, forstner2012cavity, li2018invited,indianto2021comprehensive} and actuation \cite{karunanidhi2010design, zhang2004giant}. One of their primary appeals for technological applications is that standard fabrication techniques, such as sputtering and lift-off, can be used to produce magnetostrictive thin-films. This enables scalable, low cost manufacturing of magnetostrictive devices. It is important to develop efficient and accurate methods to characterize these films.

Current techniques to measure magnetostriction in thin-films can be categorized as either direct measurements, measuring the strain induced by a magnetic field (magnetostrictive effect), or inverse, measuring the change in the magnetic properties of the material in response to an applied strain (inverse magnetostrictive effect) \cite{liang2020review}. Among direct methods, the magnetostriction of thin-films is typically measured based on the deflection of a magnetostrictively coated cantilever, using capacitive or optical techniques \cite{garcia2021magnetostrictive}. This requires careful mounting of the sample along with specialized set-ups that use AC fields, making them susceptible to vibrational noise and pick-up \cite{1989NewMagnetostrictionMethod}. To avoid these requirements, several techniques that measure the deflection under a DC field have been developed using atomic force microscopy (AFM) \cite{lima2015direct,thinfilmAFM,harin2012atomic} and nano-indentation systems \cite{shima1999accurate}. These however are single point contact measurements, which risk damage to the films and would require position scanning to obtain information about their spatial profile.

Here we present a non-contact DC method to characterize thin-film magnetostriction. The method uses optical profilometry to measure the curvature induced in a magnetostrictively coated sample by an applied DC field, from which the magnetostrictive strain is calculated using an appropriate model. This is analogous to the wafer bow method which is applied to determine the intrinsic stress of thin-films on silicon wafers \cite{ardigo2014stoney,chason2019stress,vechery2007comparison}. Our method also provides the 2D profile of the bowed film without spatial scanning, this can reveal information about the anisotropy of the film and allows for the possibility of measuring several samples simultaneously. No specialized set-up is required, relying on readily available laboratory equipment. We show that state-of-the-art sensitivity can be achieved, resolving magnetostrictive strains down to $\sim 0.1$~ppm. Moreover, the samples are compatible with other standard film characterization techniques, such as magneto-optical Kerr effect (MOKE) microscopy, x-ray diffraction (XRD) analysis, reflectometry and ellipsometry. This allows for magnetostrictive characterization along with the characterization of multiple other important material properties. 

\section{Optical Profilometry to Characterize Thin-Film Magnetostriction}

\begin{figure}[htbp]
\centering
\includegraphics[width=0.6\columnwidth]{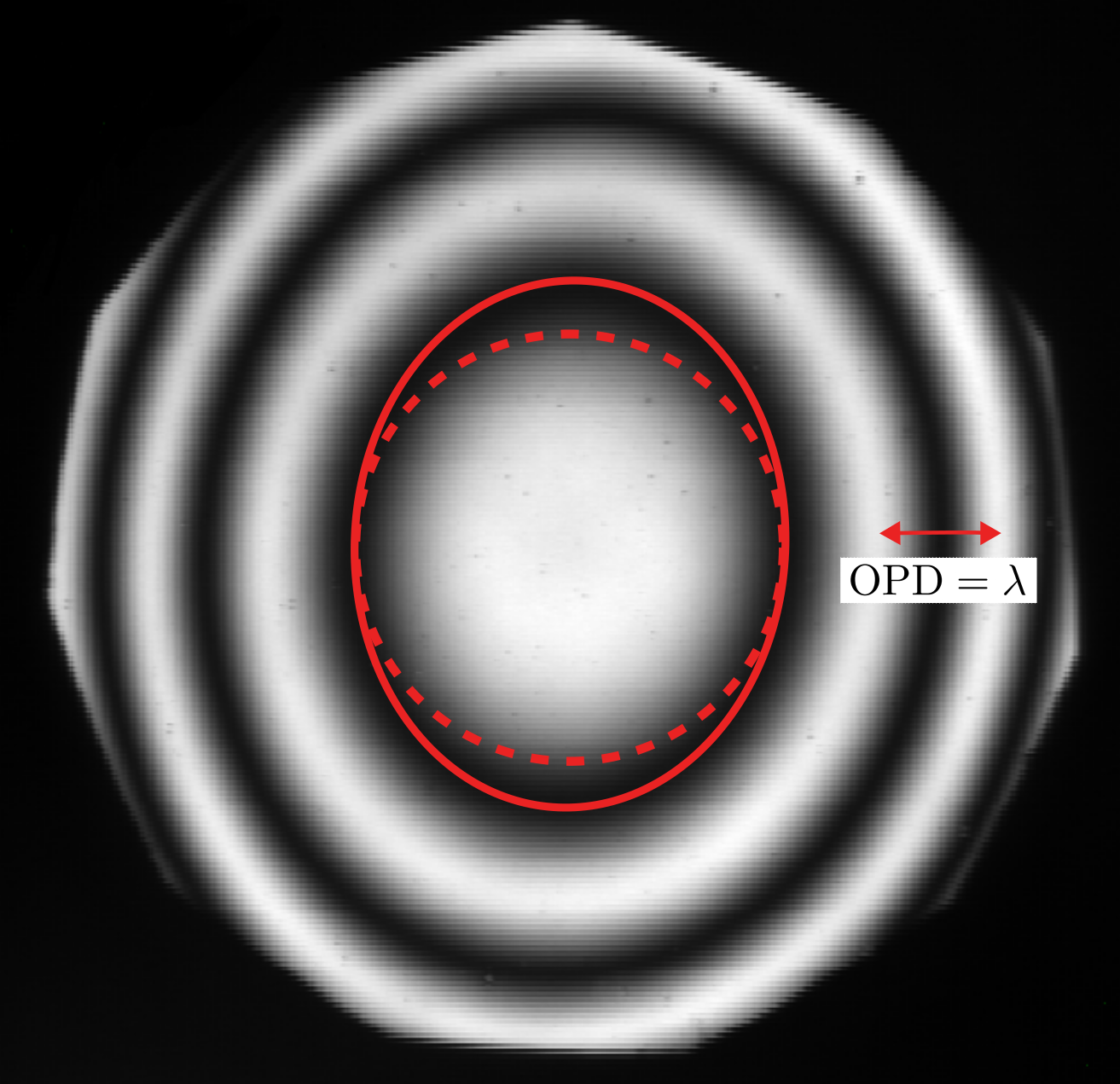}
\caption{\label{fig:966ring} Interference pattern viewed through the NT1100 profilometer, for a galfenol film sputtered on a glass coverslip, with no magnetic field applied. The interference rings are elliptical, indicating the sample is bowed anisotropically. This could be due to an anisotropic film stress, or due to anisotropy in the substrate, such as nonuniform thickness.}
\end{figure}

In our method a sample coated with a magnetostrictive thin-film is placed on a nonmagnetic stage in an optical profilometer. The sample is then imaged and a two dimensional surface profile measurement is taken, from which the intrinsic stress of the film can be calculated. A Veeco Wyko NT1100 profilometer is used, which utilizes phase-shifting interferometry (PSI), offering high resolution for mapping out smooth continuous surfaces. In the NT1100, a white-light beam is filtered to red light ($\sim$633~nm) and passed through an interferometer objective to the sample surface. Within the interferometer a beam splitter reflects half of the incident beam to a reference surface. The beams reflected from the sample and the reference recombine to form interference fringes, separated by an optical path difference (OPD) of $\lambda$. An example image of a galfenol film sputtered on a glass coverslip is shown in Fig.~(\ref{fig:966ring}), the circular rings are interference fringes due to changes in the surface profile of the film. During the measurement, a piezoelectric transducer moves the objective to cause a phase shift between the objective and sample surface. The system records the intensity of the resulting interference pattern at many different relative phase shifts, from which the surface profile is calculated \cite{creath1988v}.

\begin{figure}[htbp]
\centering
\includegraphics[width=0.95\columnwidth]{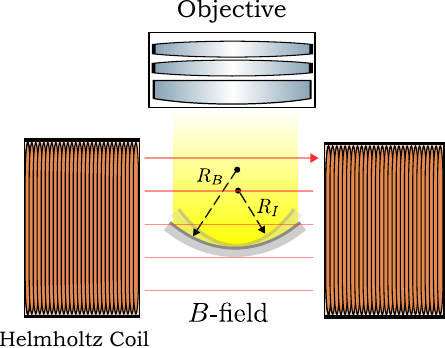}
\caption{\label{fig:profilometer} Optical profilometer experimental setup for characterizing the magnetostriction of thin-films. The surface profile of the film is measured and the change in radius of curvature is used to determine the strain induced by an applied field.}
\end{figure}

To measure the magnetostriction of a thin-film a uniform magnetic field is applied using an electromagnet, see Fig.~(\ref{fig:profilometer}). The magnetic field is aligned in-plane with the sample so that no torque is generated and the sample stays in place, with no need for clamping. The field creates a magnetostrictive strain in the material and causes the sample to deform. This deformation is observed in the profilometer, from which the strain can be determined. We use a magnetostrictive film sputtered on a circular glass coverslip. In this regime the film is much thinner than the substrate, which in turn is much smaller than the radius of the coverslip, $t_f \ll t_s \ll r_s$. Further, the deformations observed are small, with a radius of curvature much larger than the radius of the sample, $R \gg r_s $. In the case of spherical deformation, the magnetostrictive strain and stress can be calculated using Eqs.~(\ref{eq:stoneystrain}) and (\ref{eq:stoneysRR})  \cite{ardigo2014stoney, de1994magnetostriction}. Eq.~(\ref{eq:stoneysRR}) is known as Stoney's equation \cite{stoney1909tension} as applied to a circular geometry. For non-spherical deformation the extended Stoney's equations Eqs.~(\ref{eq:strainextpara}--\ref{eq:Stoneysextensionperp}) can be used. 
\begin{align}
\lambda_B &= \frac{t_s^2}{6t_f}\left(\kappa_B-\kappa_I\right) \label{eq:stoneystrain}\\
\sigma_B &=\frac{E_s}{(1-\nu_s)}\lambda_B \label{eq:stoneysRR} 
\end{align}
Here $\sigma_B$ is magnetostrictive stress induced at a given $B$-field, $E_s$ is the Young's Modulus of the substrate, $\nu_s$ is the Poisson's ratio of the substrate, $t_s$ is the thickness of the substrate, $t_f$ is the thickness of the film, $\kappa_B=\frac{1}{R_B}$ is the curvature before the $B$-field is applied, due to the intrinsic stress, and $\kappa_B$ is the curvature of the film after the $B$-field is applied, where $R$ denotes the radius of curvature \cite{ardigo2014stoney}.

\section{Deriving the Shot-Noise-Limit}

To estimate the theoretical precision of a measurement an expression for the uncertainty in the change in strain, is derived, for which any smaller change in strain could not be resolved. We present the derivation in the simplest case of isotropic deformation, however a similar derivation could be followed for any geometry and deformation. 

\subsection{Modeling the system}
In the experimentally relevant regime where radius of curvature, $R$, is much greater than the radius of the sample, the profile can be modeled by a paraboloid of the form $z=\frac{1}{2}\kappa r^2$, see Sec.~(S.1). The profilometer measures the height of the surface $z_{j,k}$ at positions $r_{j,k}$ for each pixel $(j,k)$, with noise $\delta z_{j,k}$. We assume the noise is not correlated between pixels, $\delta z_{j,k}$ is Gaussian and that the uncertainty  is the same for all pixels $\std(z_{j,k}) = \std(z)=\sqrt{\langle(\delta z)^2 \rangle}$. In Sec.~(S.3) we show that under these assumptions, for a uniform array of sufficiently dense measurements on a circular sample, the uncertainty in the optimal estimate of $\kappa$ is given by 
\begin{equation}
  \std(\kappa) \approx \sqrt{\frac{6 }{N_\mathrm{meas}}}\frac{\std(z)}{r_\mathrm{meas}^2},
    \label{eq:a_variance}
\end{equation}
where $r_{\mathrm{meas}}$ is the radius of the measurement area and $N_\mathrm{meas}$ is the number of measurement points, corresponding to the total number of pixels used to image the sample. To convert this to an uncertainty in the magnetostrictive strain we first consider the uncertainty in the change in curvature as
\begin{equation}
\small
    \std(\kappa_B- \kappa_I) = \sqrt{2}\lilspace\std( \kappa),
    \label{eq:curv_diff_variance}
\end{equation}
where we have assumed $\kappa_B$ and $ \kappa_I$ are uncorrelated and $\std( \kappa_B)=\std( \kappa_I)=\std( \kappa).$ Combining Eqs.~(\ref{eq:stoneysRR}), (\ref{eq:a_variance}) and (\ref{eq:curv_diff_variance}) gives a general expression for the uncertainty in the magnetostrictive strain in terms of $\std( z)$,
\begin{equation}
    \std(\lambda_B)= \sqrt\frac{2}{3 N_\mathrm{meas}}\frac{t_s^2}{t_f} \frac{\std( z)}{r_\mathrm{meas}^2}.
    \label{eq:stress_unc_gen}
\end{equation}
Eq.~(\ref{eq:stress_unc_gen}) is used to inform our substrate choice, and optimize the precision of a measurement, see Sec.~(\ref{sec:substratechoice}).
\subsection{Shot-noise-limit}

The precision in the case where the profilometer is operating at the shot-noise-limit can be derived to give the lower bound on the performance that can be achieved. For heterodyne optical profilometers the height of the surface is calculated from the phase change $(\phi)$ between the measurement and reference signals as 
\begin{equation}
    z(x,y)=\frac{\lambda}{4\pi} \phi (x,y)
    \label{eq:zfromphi}
\end{equation}
\cite{creath1988v}, where $\lambda$ is the wavelength of the illumination. In the shot-noise-limit, with small deviations in $\phi$ from a known mean value,  $\std(\phi)_\mathrm{shot}= \frac{1}{\sqrt{N_\mathrm{pp}}}$, where $N_\mathrm{pp}$ is the photon count per pixel \cite{hosseini2016pushing}. The uncertainty in $ z$ is therefore
\begin{equation}
   \std( z)_\mathrm{shot} =  \frac{\lambda}{4\pi} \frac{1}{\sqrt{N_\mathrm{pp}}}.
   \label{eq:stdzshot}
\end{equation}
Combining this with Eq.~(\ref{eq:stdzshot}) gives the uncertainty in the magnetostrictive strain as 
\begin{equation}
    \std(\lambda_B)=\sqrt{\frac{2}{3 N_T}}\frac{t_s^2}{t_f}\frac{\lambda}{4\pi r_\mathrm{meas}^2}.
    \label{eq:shotlimstress}
\end{equation}
Where a uniform photon count across all pixels has been assumed, with the total photon count for a given measurement taken to be  $N_T = N_\mathrm{meas} \cdot N_{pp}$. 

\subsection{Piezomagnetic coupling}

In our method we directly measure the piezomagnetic coupling as the slope of the strain vs. applied $H$-field curve $\frac{\partial \lambda_B}{\partial H}$\footnote{Note that this differs to the magnetostrictive constant which is defined along the direction of the field and at constant stress $d_{33}=\frac{\partial \lambda_{B||}}{\partial H_{||}}\big|_{\sigma}$ \cite{engdahl2000handbook}.}, where $H=B_\mathrm{applied}/\mu_0$. For many applications, such as magnetic field sensing, this is the material property which determines the performance of the device \cite{li2018invited}. To determine $\frac{\partial \lambda_B}{\partial H}$ we estimate it as linear over a small region, $\Delta H$, as
\begin{equation}
\frac{\partial \lambda_B}{\partial H}\approx \frac{\Delta \lambda_{B}}{\Delta H}.
\end{equation}
Assuming there is no uncertainty in $H$, this gives
\begin{equation}
   \std \left( \frac{\partial \lambda_B}{\partial H} \right)= \frac{\sqrt{2}\lilspace\std( \lambda_{B})}{|\Delta H|}.
\end{equation}
\section{Characterizing the magnetostriction of galfenol thin-films}
\label{characterizingmagnetostrictionfgalf}
To validate our optical profilometry based method we measured the magnetostrictive hysteresis and piezomagnetic coupling of sputtered Galfenol (FeGa) thin-films. Galfenol is an alloy of iron and gallium that has gathered significant attention in recent years owing to its high magnetostriction, excellent mechanical properties and corrosion resistance \cite{NIVEDITA2018300, Stadler}.
\subsection{Substrate choice}
\label{sec:substratechoice}
 To optimize the measurement precision and minimize $ \std(\lambda_B)$, the substrate must be  thin, compliant and large enough to fill the field of view of the profilometer, see Eq.~(\ref{eq:stress_unc_gen}). In addition, it must be able to withstand the intrinsic stresses of the films. We used 0.1~mm thick, 3~mm diameter glass coverslips, which are well suited to these requirements.  Moreover they can easily be used in other characterization techniques and are highly affordable. Other possible substrates could include free-standing membranes, which can be made as thin as 50~nm \cite{Norcada_2022}. However the model would need be adapted to optimize for this case. 

\subsection{Measurement precision}
\subsubsection{Theoretical shot-noise-limit}
To determine the shot-noise-limit of the magnetostrictive measurement given our set-up and choice of substrate, the number of photons used for a profilometer measurement must be calculated. The profilometer takes six frames of intensity data on a $60~\mathrm{fps}$ camera, from which the phase data is determined and the surface profile is calculated using Eq.~(\ref{eq:zfromphi}) \cite{creath1988v}. The illumination power used is 300~nW, corresponding to a total photon count of $N_\mathrm{meas}\approx 10^{11}$. Combining this with the system parameters $t_s=0.1~\mathrm{mm}$, $\lambda=655~\mathrm{nm}$, $r_\mathrm{meas}=0.85~\mathrm{mm}$, provides an estimate the shot-noise-limit using Eq.~(\ref{eq:shotlimstress}). For a 300~nm film this gives $\std(\lambda_B)_\mathrm{shot} \approx 0.01~\mathrm{ppm}$. In our case $\lambda_B$ can be estimated as linear over a 5~mT window for the highest slope regions, see Fig.~(\ref{fig:examplebutterfly}), corresponding to $\std \left(\frac{\partial \lambda_B}{\partial H}\right)_\mathrm{shot} \approx 4\times 10^{-3}~\mathrm{nm/A}$.

\subsubsection{Measured system noise}
\label{sec:meassystemnoise}
To measure the actual noise of the profilometer $\std(z)$, two successive profile measurements of a flat sample were taken. The difference between the profile surface heights was then calculated. The standard deviation of this difference provided $\std(z)_\mathrm{pro}\approx 0.7 \ \mathrm{nm}$,  which for the same parameters as the shot-noise calculation, corresponds to $\std(\lambda_B)_\mathrm{pro} \approx 0.1 \ \mathrm{ppm}$  and from this $\std\left(\frac{\partial \lambda_B}{\partial H}\right)_\mathrm{pro} \approx 4\times 10^{-2}~\mathrm{nm/A}$. 

This predicted sensitivity is only around a factor of ten away from the shot-noise-limit, with noise likely dominated by other noise sources such as electronic noise, environmental noise, thermal noise, and speckle noise \cite{huang1984optical,servin2009noise}. We note that the NT1100 uses a tungsten halogen lamp as a light source, which is filtered to red light. This lamp has significant intensity fluctuations compared to the LEDs used in newer models and the wavelength filter has a larger linewidth than if a laser were used, resulting in more noise. Nevertheless, the predicted strain sensitivity is on par with state-of-the-art methods which achieve strain sensitivities of $\sim$0.1~ppm using optical  \cite{1989NewMagnetostrictionMethod}, capacitive \cite{klokholm1976measurement}, nanoindentation \cite{shima1999accurate},  and inverse techniques \cite{ComparisonMagnetostrictionTechniques}.

\subsection{Sample preparation}
Magnetron DC sputtering was used to deposit Fe$_{81}$Ga$_{19}$ films, with a 3~nm Ti adhesion and a 3~nm Cu seed layer. The Ti/Cu/Fe$_{81}$Ga$_{19}$ films were sputtered in thicknesses of 206~nm, 285~nm, 419~nm, 498~nm and 966~nm onto 3~mm diameter, 0.1~mm thick borosilicate glass coverslips (Electron Microscopy Services 72296-03). The sputtering was done in a 2~mTorr argon atmosphere with a power of 150~W (DC). 

\subsection{Results}

\subsubsection{Single measurements}
\label{subsubsec:singlemeas}

\begin{figure}[htbp]
\includegraphics[width=0.95\columnwidth]{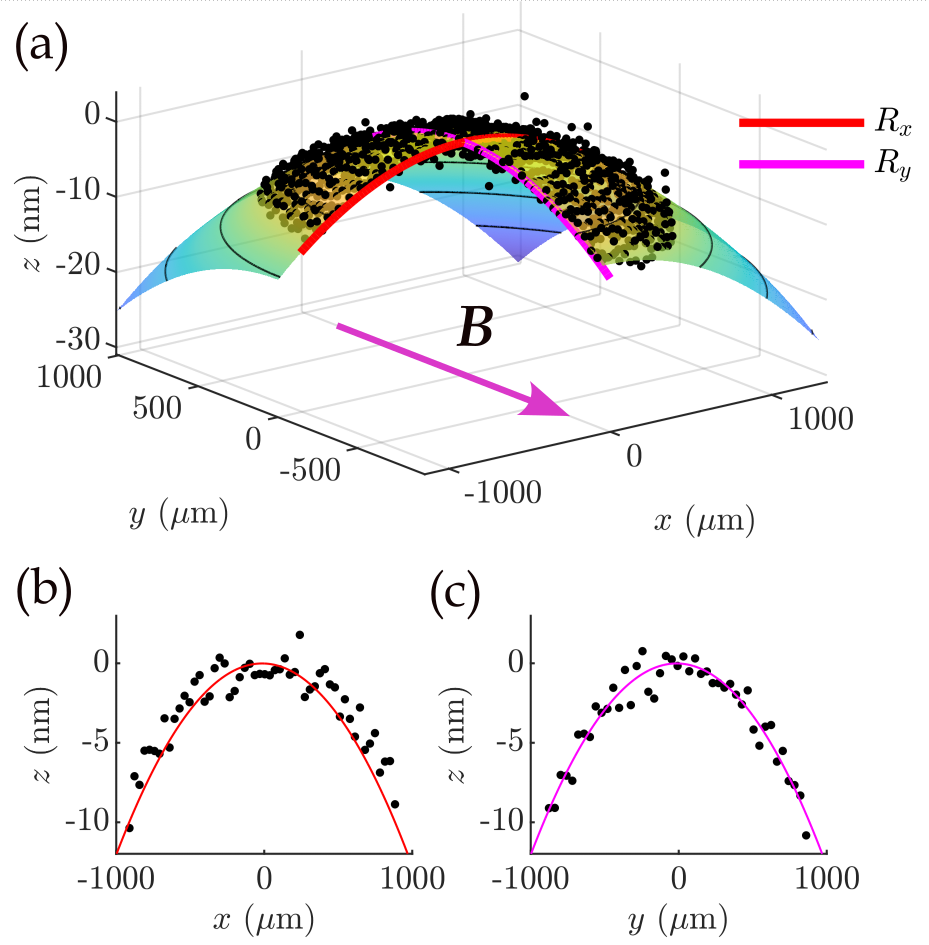}
\caption{\label{fig:subtractedsurface}
A fit of the form $z=\frac{1}{2}\kappa_x x^2+\frac{1}{2}\kappa_y y^2$ to a measurement of the deformation of a 966~nm film by a 36.5~mT field. Each point corresponds to the change in film height obtained from a given pixel. For clarity only every 1 in 10 pixels is shown such that the number of measurement points is reduced by a factor of 10. The purple arrow indicates the direction of the $B$-field. The $z=\frac{1}{2}\kappa_x x^2$ and $z=\frac{1}{2}\kappa_y y^2$ cross section curves are shown on the 3D surface in (a) and in 2D in (b) and (c).}
\end{figure}

The samples were placed on a nonmagnetic stage, 3D-printed from polylactic acid (eSUN PLA+), in the optical profilometer (Veeco Wyko NT1100), with bipole helmholtz magnetic coils (Evico Magnetics GmbH type ifw8.00.00.48) used to apply a magnetic field. To observe the deformation induced by a given $B$-field in a sample, the difference of the profiles with and without a field applied were taken and examined. The surfaces obtained from these difference measurements were not symmetric, indicating anisotropy in the magnetostrictive strain. To account for this the extension to Stoney's formula given by \cite{zhao2002evolution,sebastiani2020nano} was used, for which
\begin{align}
\lambda_{B||} &=\frac{t_s^2}{6t_f}\Delta\kappa_{B||} \label{eq:strainextpara}\\
\lambda_{B\perp} &=\frac{t_s^2}{6t_f}\Delta\kappa_{B\perp}\\
\sigma_{B||}&=\frac{E_s}{(1-\nu_s^2)}\left(\lambda_{B||} +\nu_s \lambda_{B\perp}\right) \label{eq:Stoneysextensionpara}\\
\sigma_{B\perp}&=\frac{E_s}{(1-\nu_s^2)}\left(\lambda_{B\perp} +\nu_s \lambda_{B||}\right) \label{eq:Stoneysextensionperp}
\end{align}
where $\Delta\kappa_{B||}=\kappa_{B||}-\kappa_{I||}$ and $\Delta\kappa_{B\perp}=\kappa_{B\perp}-\kappa_{I\perp}$ are the change in curvature induced in the directions parallel and perpendicular to the $B$-field. To calculate these and determine the strain an elliptic paraboloid of the form $z=\frac{1}{2}\kappa_x x^2+\frac{1}{2} \kappa_y y^2$  was fit to the difference measurements, from which we obtain $\Delta\kappa_{B\perp}=\kappa_x$ and $\Delta\kappa_{B||}=\kappa_y$, as derived in Sec.~(S.2).
A typical fit is shown in Fig.~(\ref{fig:subtractedsurface}) for the deformation of a 966~nm film by a 36.5~mT field. For this example changes in curvature of $\Delta \kappa_{B||}=-24.79\pm 0.02~\mathrm{km^{-1}}$, $\Delta \kappa_{B\perp}=-19.11\pm 0.02~\mathrm{km^{-1}}$ were measured, corresponding to magnetostrictive stresses of $\sigma_{B||} = -3.238 \pm 0.002 $~MPa, $\sigma_{B\perp}=-2.693 \pm 0.002$~MPa, and strains of $\lambda_{B||}=-42.77\pm 0.04$~ppm, $\lambda_{B\perp}=-32.97 \pm 0.03$~ppm. Here the negative sign indicates a compressive stress, due to the expansion of the film under a magnetic field. This opposes the tensile intrinsic stress of the films which is which is typically $\sim 350~\mathrm{MPa}$.

The absolute values of saturation strain $\lambda_S$ for thin-film galfenol have been previously measured in the range of 35--120~ppm \cite{NIVEDITA2018300, Basantkumar}. We did not reach saturation within our measurement range, as can be seen in Fig.~(\ref{fig:examplebutterfly}), and expect the magnetostriction value at 36.5~mT to be slightly lower but of a similar magnitude to saturation values. With this considered our result appears consistent with literature values. The strain uncertainties for this measurement are similar to, but better than, the previously reported state-of-the-art of 0.1~ppm, achieving values of $0.03$ -- $0.04~$ppm. This is close to the uncertainty of $0.02~\mathrm{ppm}$ predicted by Eq.~(\ref{eq:stress_unc_gen}) for a 966 nm film, given the measured system noise described in Sec. (\ref{sec:meassystemnoise}). The difference is likely due to imperfections in the film. In addition the optimal weighting function described in Sec. (S.3) used to derive Eq. (\ref{eq:stress_unc_gen}) has not been used, with all points weighted equally to obtain the fit in this example. 

\subsubsection{Magnetostrictive hysteresis loops}
\label{subsubsec:hysteresisloops}

\begin{figure}[htbp]
        \includegraphics[width=0.95\columnwidth]{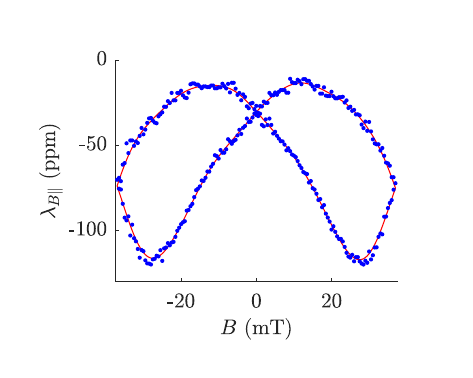}
        \caption{\label{fig:examplebutterfly} An example of a magnetostriction butterfly loop obtained for a 206~nm film. Each blue dot represents the strain along the direction of the field measured at a given $B$-field value, calculated using the method described in Sec.~(\ref{subsubsec:singlemeas}). The red trace interpolates the data, showing the `butterfly' shape of the curve. As can be seen from the shape of the loop, the magnetostriction does not saturate. This behavior was observed for all film thicknesses.}
\end{figure}

Following the single measurements the magnetostrictive hysteresis was measured. This was achieved by sweeping the $B$-field in multiple loops from $0~\mathrm{mT} \rightarrow 37~\mathrm{mT} \rightarrow -37~\mathrm{mT} \rightarrow 0~\mathrm{mT}$, in intervals of 0.5~mT, calculating the magnetostrictive strain at each point. The first loop was used to magnetize the film to a consistent magnetization for the reference profile, and the subsequent loops for measurements. As shown in Fig.~(\ref{fig:examplebutterfly}) for a 206~nm film, the magnetostriction displayed butterfly hysteresis behavior, which is characteristic of magnetostrictive materials \cite{sablik1993coupled}. 

We are able to quantify the piezomagnetic coupling of the film by the slope of these hysteresis loops. For the example in Fig.~(\ref{fig:examplebutterfly}) we obtain a maximum piezomagnetic coupling along the direction of the field of $\frac{\partial \lambda_{B||}}{\partial H}=6.0 \pm 0.3$~nm/A. For thin-film galfenol this has been measured as approximately 13~nm/A (i.e. 1.0~ppm/Oe) for varying  sputtering conditions, and thicknesses up to 480~nm  \cite{shi2019study,liang2018soft}. Hence our result is consistent with literature values, however the films could be optimized for increased performance. 

We found that the transverse magnetostriction is of the same sign as the longitudinal magnetostriction, with a slightly lower magnitude, see Fig (\ref{fig:stresscoeffs}). If Joule magnetostriction were the dominant effect it would be expected that the film undergo a longitudinal expansion and a transverse contraction \cite{de1994magnetostriction}. This suggests that the field induced changes in the elastic properties of the film, such as the magnetomechanical coupling and $\Delta E$ effect may contribute significantly to the shape and magnitude of the deformation of the sample, particularly due to the large intrinsic stress of the films.

\subsubsection{Film degradation with thickness}

\begin{figure}[htbp]
        \includegraphics[width=0.95\columnwidth]{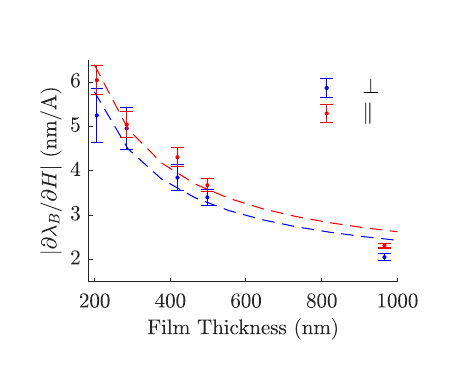}
        \caption{\label{fig:stresscoeffs} The maximum piezomagnetic coupling of the films parallel and perpendicular to the applied field versus film thickness. The fits are of the form $\big|\frac{\partial \lambda_B}{\partial H}\big|=\frac{t_1}{t}\gamma_1+ \frac{t_2}{t}\gamma_2$, where $t$ is the total thickness of the film, $t_1$, $t_2$ are the thicknesses obtained from the two-layer model described in Sec.~(S.4~B), and $\gamma_1$, $\gamma_2$ are scaling factors fit to the data, proportional to the piezomagnetic coupling of each layer. In this case $t_1=175$~nm, $t_2=x-t_1$, $\gamma_{1,y}\approx 7.0~\mathrm{nm/A}$ $\gamma_{2,y}\approx 1.7$ $\gamma_{1,x}\approx 6.4~\mathrm{nm/A}$ $\gamma_{2,x}\approx 1.6~\mathrm{nm/A}$}.
    \end{figure}

The piezomagnetic coupling decreased with film thickness as can be seen in Fig. (\ref{fig:stresscoeffs}). Degradation in magnetostriction has been previously observed, for FeGa films sputtered onto glass substrates under a 30~mT deposition field, as the thickness increased from 5~nm to 60~nm \cite{jahjah2019thickness}. This was attributed to an increase in the volume of the nonpreferential polycrystalline arrangement. Changes in the properties and structure of galfenol at larger thicknesses have also been observed. For FeGa films sputtered onto Si with a Ti/Cu seed layer \cite{AdolphiB} observed an increase in grain size and change in crystallographic texture as the thickness increased from 100~nm and 1000~nm. 

To investigate the degradation in performance in our case, we took magnetization measurements using magneto-optical kerr effect microscopy (MOKE) and vibrating sample magnetometry (VSM), see Sec.~(S.4). The magnetization hysteresis loops showed a distinct change in shape and coercivity as the samples became thicker, suggesting that the galfenol structure may enter a secondary material phase with lower magnetostriction above 206~nm. We model and present supporting data for this two-layer behavior in Sec.~(S.4~B). 

\section{\label{sec:discussion} Conclusion}

We have presented a new method to measure the magnetostriction of thin-films with state-of-the-art sensitivity using optical profilometry. This method offers benefits over common techniques such as AC field driven cantilever methods. These include that it is non-contact, allows for magnetostriction measurements under a DC field and grants information about the anisotropy of the films. This method could be extended to other smaller substrates such as free-standing membranes, which would allow for multiple substrates to be arrayed and measured simultaneously. Alternatively, larger substrates could be used in combination with a wider field of view objective to further improve the sensitivity. 

\begin{acknowledgments}
The authors acknowledge the facilities, and the scientific and technical assistance, of the Australian Microscopy \& Microanalysis Research Facility at the Centre for Microscopy and Microanalysis, The University of Queensland. This work was performed in part at the Queensland node of the Australian National Fabrication Facility. A company established under the National Collaborative Research Infrastructure Strategy to provide nano and microfabrication facilities for Australia's researchers. The Authors also acknowledge the highly valuable advice and support provided by Rodney Appleby. This work was supported by the Australian Government through the Next Generation Technologies Fund (now managed through ASCA). This work was financially supported by the Australian Research Council (ARC) Centre of excellence for Engineered Quantum systems (EQUS): Grant No. CE170100009 and Orica Limited.
\end{acknowledgments}

\bibliography{apssamp.bib}

\providecommand{\noopsort}[1]{}\providecommand{\singleletter}[1]{#1}%
\begin{thebibliography}{43}%
\makeatletter
\providecommand \@ifxundefined [1]{%
 \@ifx{#1\undefined}
}%
\providecommand \@ifnum [1]{%
 \ifnum #1\expandafter \@firstoftwo
 \else \expandafter \@secondoftwo
 \fi
}%
\providecommand \@ifx [1]{%
 \ifx #1\expandafter \@firstoftwo
 \else \expandafter \@secondoftwo
 \fi
}%
\providecommand \natexlab [1]{#1}%
\providecommand \enquote  [1]{``#1''}%
\providecommand \bibnamefont  [1]{#1}%
\providecommand \bibfnamefont [1]{#1}%
\providecommand \citenamefont [1]{#1}%
\providecommand \href@noop [0]{\@secondoftwo}%
\providecommand \href [0]{\begingroup \@sanitize@url \@href}%
\providecommand \@href[1]{\@@startlink{#1}\@@href}%
\providecommand \@@href[1]{\endgroup#1\@@endlink}%
\providecommand \@sanitize@url [0]{\catcode `\\12\catcode `\$12\catcode `\&12\catcode `\#12\catcode `\^12\catcode `\_12\catcode `\%12\relax}%
\providecommand \@@startlink[1]{}%
\providecommand \@@endlink[0]{}%
\providecommand \url  [0]{\begingroup\@sanitize@url \@url }%
\providecommand \@url [1]{\endgroup\@href {#1}{\urlprefix }}%
\providecommand \urlprefix  [0]{URL }%
\providecommand \Eprint [0]{\href }%
\providecommand \doibase [0]{https://doi.org/}%
\providecommand \selectlanguage [0]{\@gobble}%
\providecommand \bibinfo  [0]{\@secondoftwo}%
\providecommand \bibfield  [0]{\@secondoftwo}%
\providecommand \translation [1]{[#1]}%
\providecommand \BibitemOpen [0]{}%
\providecommand \bibitemStop [0]{}%
\providecommand \bibitemNoStop [0]{.\EOS\space}%
\providecommand \EOS [0]{\spacefactor3000\relax}%
\providecommand \BibitemShut  [1]{\csname bibitem#1\endcsname}%
\let\auto@bib@innerbib\@empty
\bibitem [{\citenamefont {Wang}\ \emph {et~al.}(2023)\citenamefont {Wang}, \citenamefont {Xiang}, \citenamefont {Yu},\ and\ \citenamefont {Yang}}]{wang2023development}%
  \BibitemOpen
  \bibfield  {author} {\bibinfo {author} {\bibfnamefont {W.}~\bibnamefont {Wang}}, \bibinfo {author} {\bibfnamefont {Y.}~\bibnamefont {Xiang}}, \bibinfo {author} {\bibfnamefont {J.}~\bibnamefont {Yu}},\ and\ \bibinfo {author} {\bibfnamefont {L.}~\bibnamefont {Yang}},\ }\bibfield  {title} {\bibinfo {title} {Development and prospect of smart materials and structures for aerospace sensing systems and applications},\ }\href@noop {} {\bibfield  {journal} {\bibinfo  {journal} {Sensors}\ }\textbf {\bibinfo {volume} {23}},\ \bibinfo {pages} {1545} (\bibinfo {year} {2023})}\BibitemShut {NoStop}%
\bibitem [{\citenamefont {Zanjanchi}\ \emph {et~al.}(2023)\citenamefont {Zanjanchi}, \citenamefont {Ghadiri}, \citenamefont {Sabouri-Ghomi},\ and\ \citenamefont {Mirzaghafoor}}]{zanjanchi2023bifurcation}%
  \BibitemOpen
  \bibfield  {author} {\bibinfo {author} {\bibfnamefont {M.}~\bibnamefont {Zanjanchi}}, \bibinfo {author} {\bibfnamefont {M.}~\bibnamefont {Ghadiri}}, \bibinfo {author} {\bibfnamefont {S.}~\bibnamefont {Sabouri-Ghomi}},\ and\ \bibinfo {author} {\bibfnamefont {K.}~\bibnamefont {Mirzaghafoor}},\ }\bibfield  {title} {\bibinfo {title} {Bifurcation point analysis of a magnetostrictive sandwich composite plate subjected to magnetic field and axial force},\ }\href@noop {} {\bibfield  {journal} {\bibinfo  {journal} {International Journal of Structural Stability and Dynamics}\ ,\ \bibinfo {pages} {2350172}} (\bibinfo {year} {2023})}\BibitemShut {NoStop}%
\bibitem [{\citenamefont {Lu}\ \emph {et~al.}(2023)\citenamefont {Lu}, \citenamefont {Zhou}, \citenamefont {Wu}, \citenamefont {Xiao}, \citenamefont {Zhang}, \citenamefont {Wang}, \citenamefont {He}, \citenamefont {Yang},\ and\ \citenamefont {Fu}}]{lu2023cylindrical}%
  \BibitemOpen
  \bibfield  {author} {\bibinfo {author} {\bibfnamefont {C.}~\bibnamefont {Lu}}, \bibinfo {author} {\bibfnamefont {H.}~\bibnamefont {Zhou}}, \bibinfo {author} {\bibfnamefont {G.}~\bibnamefont {Wu}}, \bibinfo {author} {\bibfnamefont {X.}~\bibnamefont {Xiao}}, \bibinfo {author} {\bibfnamefont {Z.}~\bibnamefont {Zhang}}, \bibinfo {author} {\bibfnamefont {J.}~\bibnamefont {Wang}}, \bibinfo {author} {\bibfnamefont {X.}~\bibnamefont {He}}, \bibinfo {author} {\bibfnamefont {A.}~\bibnamefont {Yang}},\ and\ \bibinfo {author} {\bibfnamefont {G.}~\bibnamefont {Fu}},\ }\bibfield  {title} {\bibinfo {title} {Cylindrical magnetoelectric fega/pzt composite for lightning current sensing applications},\ }\href@noop {} {\bibfield  {journal} {\bibinfo  {journal} {Smart Materials and Structures}\ } (\bibinfo {year} {2023})}\BibitemShut {NoStop}%
\bibitem [{\citenamefont {Gotardo}\ \emph {et~al.}(2023)\citenamefont {Gotardo}, \citenamefont {Carey}, \citenamefont {Greenall}, \citenamefont {Harris}, \citenamefont {Romero}, \citenamefont {Bulla}, \citenamefont {Bridge}, \citenamefont {Bennett}, \citenamefont {Foster},\ and\ \citenamefont {Bowen}}]{gotardo2023waveguide}%
  \BibitemOpen
  \bibfield  {author} {\bibinfo {author} {\bibfnamefont {F.}~\bibnamefont {Gotardo}}, \bibinfo {author} {\bibfnamefont {B.~J.}\ \bibnamefont {Carey}}, \bibinfo {author} {\bibfnamefont {H.}~\bibnamefont {Greenall}}, \bibinfo {author} {\bibfnamefont {G.~I.}\ \bibnamefont {Harris}}, \bibinfo {author} {\bibfnamefont {E.}~\bibnamefont {Romero}}, \bibinfo {author} {\bibfnamefont {D.}~\bibnamefont {Bulla}}, \bibinfo {author} {\bibfnamefont {E.~M.}\ \bibnamefont {Bridge}}, \bibinfo {author} {\bibfnamefont {J.~S.}\ \bibnamefont {Bennett}}, \bibinfo {author} {\bibfnamefont {S.}~\bibnamefont {Foster}},\ and\ \bibinfo {author} {\bibfnamefont {W.~P.}\ \bibnamefont {Bowen}},\ }\bibfield  {title} {\bibinfo {title} {Waveguide-integrated and portable optomechanical magnetometer},\ }\href@noop {} {\bibfield  {journal} {\bibinfo  {journal} {arXiv preprint arXiv:2307.15229}\ } (\bibinfo {year} {2023})}\BibitemShut {NoStop}%
\bibitem [{\citenamefont {Li}\ \emph {et~al.}(2012)\citenamefont {Li}, \citenamefont {Dhagat},\ and\ \citenamefont {Jander}}]{li2012surface}%
  \BibitemOpen
  \bibfield  {author} {\bibinfo {author} {\bibfnamefont {W.}~\bibnamefont {Li}}, \bibinfo {author} {\bibfnamefont {P.}~\bibnamefont {Dhagat}},\ and\ \bibinfo {author} {\bibfnamefont {A.}~\bibnamefont {Jander}},\ }\bibfield  {title} {\bibinfo {title} {Surface acoustic wave magnetic sensor using galfenol thin film},\ }\href@noop {} {\bibfield  {journal} {\bibinfo  {journal} {IEEE transactions on magnetics}\ }\textbf {\bibinfo {volume} {48}},\ \bibinfo {pages} {4100} (\bibinfo {year} {2012})}\BibitemShut {NoStop}%
\bibitem [{\citenamefont {Forstner}\ \emph {et~al.}(2012)\citenamefont {Forstner}, \citenamefont {Prams}, \citenamefont {Knittel}, \citenamefont {Van~Ooijen}, \citenamefont {Swaim}, \citenamefont {Harris}, \citenamefont {Szorkovszky}, \citenamefont {Bowen},\ and\ \citenamefont {Rubinsztein-Dunlop}}]{forstner2012cavity}%
  \BibitemOpen
  \bibfield  {author} {\bibinfo {author} {\bibfnamefont {S.}~\bibnamefont {Forstner}}, \bibinfo {author} {\bibfnamefont {S.}~\bibnamefont {Prams}}, \bibinfo {author} {\bibfnamefont {J.}~\bibnamefont {Knittel}}, \bibinfo {author} {\bibfnamefont {E.}~\bibnamefont {Van~Ooijen}}, \bibinfo {author} {\bibfnamefont {J.}~\bibnamefont {Swaim}}, \bibinfo {author} {\bibfnamefont {G.}~\bibnamefont {Harris}}, \bibinfo {author} {\bibfnamefont {A.}~\bibnamefont {Szorkovszky}}, \bibinfo {author} {\bibfnamefont {W.}~\bibnamefont {Bowen}},\ and\ \bibinfo {author} {\bibfnamefont {H.}~\bibnamefont {Rubinsztein-Dunlop}},\ }\bibfield  {title} {\bibinfo {title} {Cavity optomechanical magnetometer},\ }\href@noop {} {\bibfield  {journal} {\bibinfo  {journal} {Physical review letters}\ }\textbf {\bibinfo {volume} {108}},\ \bibinfo {pages} {120801} (\bibinfo {year} {2012})}\BibitemShut {NoStop}%
\bibitem [{\citenamefont {Li}\ \emph {et~al.}(2018)\citenamefont {Li}, \citenamefont {Bulla}, \citenamefont {Prakash}, \citenamefont {Forstner}, \citenamefont {Dehghan-Manshadi}, \citenamefont {Rubinsztein-Dunlop}, \citenamefont {Foster},\ and\ \citenamefont {Bowen}}]{li2018invited}%
  \BibitemOpen
  \bibfield  {author} {\bibinfo {author} {\bibfnamefont {B.-B.}\ \bibnamefont {Li}}, \bibinfo {author} {\bibfnamefont {D.}~\bibnamefont {Bulla}}, \bibinfo {author} {\bibfnamefont {V.}~\bibnamefont {Prakash}}, \bibinfo {author} {\bibfnamefont {S.}~\bibnamefont {Forstner}}, \bibinfo {author} {\bibfnamefont {A.}~\bibnamefont {Dehghan-Manshadi}}, \bibinfo {author} {\bibfnamefont {H.}~\bibnamefont {Rubinsztein-Dunlop}}, \bibinfo {author} {\bibfnamefont {S.}~\bibnamefont {Foster}},\ and\ \bibinfo {author} {\bibfnamefont {W.~P.}\ \bibnamefont {Bowen}},\ }\bibfield  {title} {\bibinfo {title} {Invited article: Scalable high-sensitivity optomechanical magnetometers on a chip},\ }\href@noop {} {\bibfield  {journal} {\bibinfo  {journal} {Apl Photonics}\ }\textbf {\bibinfo {volume} {3}} (\bibinfo {year} {2018})}\BibitemShut {NoStop}%
\bibitem [{\citenamefont {Indianto}\ \emph {et~al.}(2021)\citenamefont {Indianto}, \citenamefont {Toda},\ and\ \citenamefont {Ono}}]{indianto2021comprehensive}%
  \BibitemOpen
  \bibfield  {author} {\bibinfo {author} {\bibfnamefont {M.~A.}\ \bibnamefont {Indianto}}, \bibinfo {author} {\bibfnamefont {M.}~\bibnamefont {Toda}},\ and\ \bibinfo {author} {\bibfnamefont {T.}~\bibnamefont {Ono}},\ }\bibfield  {title} {\bibinfo {title} {Comprehensive study of magnetostriction-based mems magnetic sensor of a fega/pzt cantilever},\ }\href@noop {} {\bibfield  {journal} {\bibinfo  {journal} {Sensors and Actuators A: Physical}\ }\textbf {\bibinfo {volume} {331}},\ \bibinfo {pages} {112985} (\bibinfo {year} {2021})}\BibitemShut {NoStop}%
\bibitem [{\citenamefont {Karunanidhi}\ and\ \citenamefont {Singaperumal}(2010)}]{karunanidhi2010design}%
  \BibitemOpen
  \bibfield  {author} {\bibinfo {author} {\bibfnamefont {S.}~\bibnamefont {Karunanidhi}}\ and\ \bibinfo {author} {\bibfnamefont {M.}~\bibnamefont {Singaperumal}},\ }\bibfield  {title} {\bibinfo {title} {Design, analysis and simulation of magnetostrictive actuator and its application to high dynamic servo valve},\ }\href@noop {} {\bibfield  {journal} {\bibinfo  {journal} {Sensors and Actuators A: Physical}\ }\textbf {\bibinfo {volume} {157}},\ \bibinfo {pages} {185} (\bibinfo {year} {2010})}\BibitemShut {NoStop}%
\bibitem [{\citenamefont {Zhang}\ \emph {et~al.}(2004)\citenamefont {Zhang}, \citenamefont {Jiang}, \citenamefont {Zhang},\ and\ \citenamefont {Xu}}]{zhang2004giant}%
  \BibitemOpen
  \bibfield  {author} {\bibinfo {author} {\bibfnamefont {T.}~\bibnamefont {Zhang}}, \bibinfo {author} {\bibfnamefont {C.}~\bibnamefont {Jiang}}, \bibinfo {author} {\bibfnamefont {H.}~\bibnamefont {Zhang}},\ and\ \bibinfo {author} {\bibfnamefont {H.}~\bibnamefont {Xu}},\ }\bibfield  {title} {\bibinfo {title} {Giant magnetostrictive actuators for active vibration control},\ }\href@noop {} {\bibfield  {journal} {\bibinfo  {journal} {Smart materials and structures}\ }\textbf {\bibinfo {volume} {13}},\ \bibinfo {pages} {473} (\bibinfo {year} {2004})}\BibitemShut {NoStop}%
\bibitem [{\citenamefont {Liang}\ \emph {et~al.}(2020)\citenamefont {Liang}, \citenamefont {Dong}, \citenamefont {Chen}, \citenamefont {Wang}, \citenamefont {Wei}, \citenamefont {Zaeimbashi}, \citenamefont {He}, \citenamefont {Matyushov}, \citenamefont {Sun},\ and\ \citenamefont {Sun}}]{liang2020review}%
  \BibitemOpen
  \bibfield  {author} {\bibinfo {author} {\bibfnamefont {X.}~\bibnamefont {Liang}}, \bibinfo {author} {\bibfnamefont {C.}~\bibnamefont {Dong}}, \bibinfo {author} {\bibfnamefont {H.}~\bibnamefont {Chen}}, \bibinfo {author} {\bibfnamefont {J.}~\bibnamefont {Wang}}, \bibinfo {author} {\bibfnamefont {Y.}~\bibnamefont {Wei}}, \bibinfo {author} {\bibfnamefont {M.}~\bibnamefont {Zaeimbashi}}, \bibinfo {author} {\bibfnamefont {Y.}~\bibnamefont {He}}, \bibinfo {author} {\bibfnamefont {A.}~\bibnamefont {Matyushov}}, \bibinfo {author} {\bibfnamefont {C.}~\bibnamefont {Sun}},\ and\ \bibinfo {author} {\bibfnamefont {N.}~\bibnamefont {Sun}},\ }\bibfield  {title} {\bibinfo {title} {A review of thin-film magnetoelastic materials for magnetoelectric applications},\ }\href@noop {} {\bibfield  {journal} {\bibinfo  {journal} {Sensors}\ }\textbf {\bibinfo {volume} {20}},\ \bibinfo {pages} {1532} (\bibinfo {year} {2020})}\BibitemShut {NoStop}%
\bibitem [{\citenamefont {Garc{\'\i}a-Arribas}(2021)}]{garcia2021magnetostrictive}%
  \BibitemOpen
  \bibfield  {author} {\bibinfo {author} {\bibfnamefont {A.}~\bibnamefont {Garc{\'\i}a-Arribas}},\ }\bibfield  {title} {\bibinfo {title} {Magnetostrictive materials},\ }\href@noop {} {\bibfield  {journal} {\bibinfo  {journal} {Magnetic Measurement Techniques for Materials Characterization}\ ,\ \bibinfo {pages} {727}} (\bibinfo {year} {2021})}\BibitemShut {NoStop}%
\bibitem [{\citenamefont {Tam}\ and\ \citenamefont {Schroeder}(1989)}]{1989NewMagnetostrictionMethod}%
  \BibitemOpen
  \bibfield  {author} {\bibinfo {author} {\bibfnamefont {A.}~\bibnamefont {Tam}}\ and\ \bibinfo {author} {\bibfnamefont {H.}~\bibnamefont {Schroeder}},\ }\bibfield  {title} {\bibinfo {title} {A new high-precision optical technique to measure magnetostriction of a thin magnetic film deposited on a substrate},\ }\href {https://doi.org/10.1109/20.24502} {\bibfield  {journal} {\bibinfo  {journal} {IEEE Transactions on Magnetics}\ }\textbf {\bibinfo {volume} {25}},\ \bibinfo {pages} {2629} (\bibinfo {year} {1989})}\BibitemShut {NoStop}%
\bibitem [{\citenamefont {Lima}\ \emph {et~al.}(2015)\citenamefont {Lima}, \citenamefont {Maximino}, \citenamefont {Santos},\ and\ \citenamefont {Santos}}]{lima2015direct}%
  \BibitemOpen
  \bibfield  {author} {\bibinfo {author} {\bibfnamefont {B.~L. S.~d.}\ \bibnamefont {Lima}}, \bibinfo {author} {\bibfnamefont {F.~L.}\ \bibnamefont {Maximino}}, \bibinfo {author} {\bibfnamefont {J.~C.}\ \bibnamefont {Santos}},\ and\ \bibinfo {author} {\bibfnamefont {A.}~\bibnamefont {Santos}},\ }\bibfield  {title} {\bibinfo {title} {Direct method for magnetostriction coefficient measurement based on atomic force microscope, illustrated by the example of tb--co film},\ }\href@noop {} {\bibfield  {journal} {\bibinfo  {journal} {Journal of Magnetism and Magnetic Materials}\ }\textbf {\bibinfo {volume} {395}},\ \bibinfo {pages} {336} (\bibinfo {year} {2015})}\BibitemShut {NoStop}%
\bibitem [{\citenamefont {Coïsson}\ \emph {et~al.}(2020)\citenamefont {Coïsson}, \citenamefont {Hüttenes}, \citenamefont {Cialone}, \citenamefont {Barrera}, \citenamefont {Celegato}, \citenamefont {Rizzi}, \citenamefont {Barber},\ and\ \citenamefont {Tiberto}}]{thinfilmAFM}%
  \BibitemOpen
  \bibfield  {author} {\bibinfo {author} {\bibfnamefont {M.}~\bibnamefont {Coïsson}}, \bibinfo {author} {\bibfnamefont {W.}~\bibnamefont {Hüttenes}}, \bibinfo {author} {\bibfnamefont {M.}~\bibnamefont {Cialone}}, \bibinfo {author} {\bibfnamefont {G.}~\bibnamefont {Barrera}}, \bibinfo {author} {\bibfnamefont {F.}~\bibnamefont {Celegato}}, \bibinfo {author} {\bibfnamefont {P.}~\bibnamefont {Rizzi}}, \bibinfo {author} {\bibfnamefont {Z.~H.}\ \bibnamefont {Barber}},\ and\ \bibinfo {author} {\bibfnamefont {P.}~\bibnamefont {Tiberto}},\ }\bibfield  {title} {\bibinfo {title} {Measurement of thin film magnetostriction using field-dependent atomic force microscopy},\ }\href {https://doi.org/https://doi.org/10.1016/j.apsusc.2020.146514} {\bibfield  {journal} {\bibinfo  {journal} {Applied Surface Science}\ }\textbf {\bibinfo {volume} {525}},\ \bibinfo {pages} {146514} (\bibinfo {year} {2020})}\BibitemShut {NoStop}%
\bibitem [{\citenamefont {Harin}\ \emph {et~al.}(2012)\citenamefont {Harin}, \citenamefont {Sheftel’},\ and\ \citenamefont {Krikunov}}]{harin2012atomic}%
  \BibitemOpen
  \bibfield  {author} {\bibinfo {author} {\bibfnamefont {E.}~\bibnamefont {Harin}}, \bibinfo {author} {\bibfnamefont {E.}~\bibnamefont {Sheftel’}},\ and\ \bibinfo {author} {\bibfnamefont {A.}~\bibnamefont {Krikunov}},\ }\bibfield  {title} {\bibinfo {title} {Atomic force microscopy measurements of magnetostriction of soft-magnetic films},\ }\href@noop {} {\bibfield  {journal} {\bibinfo  {journal} {Solid State Phenomena}\ }\textbf {\bibinfo {volume} {190}},\ \bibinfo {pages} {179} (\bibinfo {year} {2012})}\BibitemShut {NoStop}%
\bibitem [{\citenamefont {Shima}\ and\ \citenamefont {Fujimori}(1999)}]{shima1999accurate}%
  \BibitemOpen
  \bibfield  {author} {\bibinfo {author} {\bibfnamefont {T.}~\bibnamefont {Shima}}\ and\ \bibinfo {author} {\bibfnamefont {H.}~\bibnamefont {Fujimori}},\ }\bibfield  {title} {\bibinfo {title} {An accurate measurement of magnetostriction of thin films by using nano-indentation system},\ }\href@noop {} {\bibfield  {journal} {\bibinfo  {journal} {IEEE Transactions on magnetics}\ }\textbf {\bibinfo {volume} {35}},\ \bibinfo {pages} {3832} (\bibinfo {year} {1999})}\BibitemShut {NoStop}%
\bibitem [{\citenamefont {Ardigo}\ \emph {et~al.}(2014)\citenamefont {Ardigo}, \citenamefont {Ahmed},\ and\ \citenamefont {Besnard}}]{ardigo2014stoney}%
  \BibitemOpen
  \bibfield  {author} {\bibinfo {author} {\bibfnamefont {M.~R.}\ \bibnamefont {Ardigo}}, \bibinfo {author} {\bibfnamefont {M.}~\bibnamefont {Ahmed}},\ and\ \bibinfo {author} {\bibfnamefont {A.}~\bibnamefont {Besnard}},\ }\bibfield  {title} {\bibinfo {title} {Stoney formula: Investigation of curvature measurements by optical profilometer},\ }\href@noop {} {\bibfield  {journal} {\bibinfo  {journal} {Advanced Materials Research}\ }\textbf {\bibinfo {volume} {996}},\ \bibinfo {pages} {361} (\bibinfo {year} {2014})}\BibitemShut {NoStop}%
\bibitem [{\citenamefont {Chason}(2019)}]{chason2019stress}%
  \BibitemOpen
  \bibfield  {author} {\bibinfo {author} {\bibfnamefont {E.}~\bibnamefont {Chason}},\ }\bibfield  {title} {\bibinfo {title} {Stress measurement in thin films using wafer curvature: Principles and applications},\ }in\ \href@noop {} {\emph {\bibinfo {booktitle} {Handbook of Mechanics of Materials}}}\ (\bibinfo  {publisher} {Springer},\ \bibinfo {year} {2019})\ pp.\ \bibinfo {pages} {2051--2082}\BibitemShut {NoStop}%
\bibitem [{\citenamefont {Vechery}\ \emph {et~al.}(2007)\citenamefont {Vechery}, \citenamefont {Dick}, \citenamefont {Balachandran},\ and\ \citenamefont {Dubey}}]{vechery2007comparison}%
  \BibitemOpen
  \bibfield  {author} {\bibinfo {author} {\bibfnamefont {M.}~\bibnamefont {Vechery}}, \bibinfo {author} {\bibfnamefont {A.}~\bibnamefont {Dick}}, \bibinfo {author} {\bibfnamefont {B.}~\bibnamefont {Balachandran}},\ and\ \bibinfo {author} {\bibfnamefont {M.}~\bibnamefont {Dubey}},\ }\bibfield  {title} {\bibinfo {title} {Comparison of techniques for measurement of residual stresses in multilayered micro-electro-mechanical devices},\ }in\ \href@noop {} {\emph {\bibinfo {booktitle} {Nanosensors, Microsensors, and Biosensors and Systems 2007}}},\ Vol.\ \bibinfo {volume} {6528}\ (\bibinfo {organization} {SPIE},\ \bibinfo {year} {2007})\ pp.\ \bibinfo {pages} {99--110}\BibitemShut {NoStop}%
\bibitem [{\citenamefont {Creath}(1988)}]{creath1988v}%
  \BibitemOpen
  \bibfield  {author} {\bibinfo {author} {\bibfnamefont {K.}~\bibnamefont {Creath}},\ }\bibfield  {title} {\bibinfo {title} {V phase-measurement interferometry techniques},\ }in\ \href@noop {} {\emph {\bibinfo {booktitle} {Progress in optics}}},\ Vol.~\bibinfo {volume} {26}\ (\bibinfo  {publisher} {Elsevier},\ \bibinfo {year} {1988})\ pp.\ \bibinfo {pages} {349--393}\BibitemShut {NoStop}%
\bibitem [{\citenamefont {De~Lacheisserie}\ and\ \citenamefont {Peuzin}(1994)}]{de1994magnetostriction}%
  \BibitemOpen
  \bibfield  {author} {\bibinfo {author} {\bibfnamefont {E.~d.~T.}\ \bibnamefont {De~Lacheisserie}}\ and\ \bibinfo {author} {\bibfnamefont {J.}~\bibnamefont {Peuzin}},\ }\bibfield  {title} {\bibinfo {title} {Magnetostriction and internal stresses in thin films: the cantilever method revisited},\ }\href@noop {} {\bibfield  {journal} {\bibinfo  {journal} {Journal of Magnetism and Magnetic Materials}\ }\textbf {\bibinfo {volume} {136}},\ \bibinfo {pages} {189} (\bibinfo {year} {1994})}\BibitemShut {NoStop}%
\bibitem [{\citenamefont {Stoney}(1909)}]{stoney1909tension}%
  \BibitemOpen
  \bibfield  {author} {\bibinfo {author} {\bibfnamefont {G.~G.}\ \bibnamefont {Stoney}},\ }\bibfield  {title} {\bibinfo {title} {The tension of metallic films deposited by electrolysis},\ }\href@noop {} {\bibfield  {journal} {\bibinfo  {journal} {Proceedings of the Royal Society of London. Series A, Containing Papers of a Mathematical and Physical Character}\ }\textbf {\bibinfo {volume} {82}},\ \bibinfo {pages} {172} (\bibinfo {year} {1909})}\BibitemShut {NoStop}%
\bibitem [{\citenamefont {Hosseini}\ \emph {et~al.}(2016)\citenamefont {Hosseini}, \citenamefont {Zhou}, \citenamefont {Kim}, \citenamefont {Peres}, \citenamefont {Diaspro}, \citenamefont {Kuang}, \citenamefont {Yaqoob},\ and\ \citenamefont {So}}]{hosseini2016pushing}%
  \BibitemOpen
  \bibfield  {author} {\bibinfo {author} {\bibfnamefont {P.}~\bibnamefont {Hosseini}}, \bibinfo {author} {\bibfnamefont {R.}~\bibnamefont {Zhou}}, \bibinfo {author} {\bibfnamefont {Y.-H.}\ \bibnamefont {Kim}}, \bibinfo {author} {\bibfnamefont {C.}~\bibnamefont {Peres}}, \bibinfo {author} {\bibfnamefont {A.}~\bibnamefont {Diaspro}}, \bibinfo {author} {\bibfnamefont {C.}~\bibnamefont {Kuang}}, \bibinfo {author} {\bibfnamefont {Z.}~\bibnamefont {Yaqoob}},\ and\ \bibinfo {author} {\bibfnamefont {P.~T.}\ \bibnamefont {So}},\ }\bibfield  {title} {\bibinfo {title} {Pushing phase and amplitude sensitivity limits in interferometric microscopy},\ }\href@noop {} {\bibfield  {journal} {\bibinfo  {journal} {Optics letters}\ }\textbf {\bibinfo {volume} {41}},\ \bibinfo {pages} {1656} (\bibinfo {year} {2016})}\BibitemShut {NoStop}%
\bibitem [{\citenamefont {Engdahl}\ and\ \citenamefont {Mayergoyz}(2000)}]{engdahl2000handbook}%
  \BibitemOpen
  \bibfield  {author} {\bibinfo {author} {\bibfnamefont {G.}~\bibnamefont {Engdahl}}\ and\ \bibinfo {author} {\bibfnamefont {I.~D.}\ \bibnamefont {Mayergoyz}},\ }\href@noop {} {\emph {\bibinfo {title} {Handbook of giant magnetostrictive materials}}},\ Vol.\ \bibinfo {volume} {386}\ (\bibinfo  {publisher} {Elsevier},\ \bibinfo {year} {2000})\BibitemShut {NoStop}%
\bibitem [{\citenamefont {Nivedita}\ \emph {et~al.}(2018)\citenamefont {Nivedita}, \citenamefont {Manivel}, \citenamefont {Pandian}, \citenamefont {Murugesan}, \citenamefont {Morley}, \citenamefont {Asokan},\ and\ \citenamefont {{Rajendra Kumar}}}]{NIVEDITA2018300}%
  \BibitemOpen
  \bibfield  {author} {\bibinfo {author} {\bibfnamefont {L.~R.}\ \bibnamefont {Nivedita}}, \bibinfo {author} {\bibfnamefont {P.}~\bibnamefont {Manivel}}, \bibinfo {author} {\bibfnamefont {R.}~\bibnamefont {Pandian}}, \bibinfo {author} {\bibfnamefont {S.}~\bibnamefont {Murugesan}}, \bibinfo {author} {\bibfnamefont {N.~A.}\ \bibnamefont {Morley}}, \bibinfo {author} {\bibfnamefont {K.}~\bibnamefont {Asokan}},\ and\ \bibinfo {author} {\bibfnamefont {R.~T.}\ \bibnamefont {{Rajendra Kumar}}},\ }\bibfield  {title} {\bibinfo {title} {Enhancement of magnetostrictive properties of galfenol thin films},\ }\href {https://doi.org/https://doi.org/10.1016/j.jmmm.2017.11.030} {\bibfield  {journal} {\bibinfo  {journal} {Journal of Magnetism and Magnetic Materials}\ }\textbf {\bibinfo {volume} {451}},\ \bibinfo {pages} {300} (\bibinfo {year} {2018})}\BibitemShut {NoStop}%
\bibitem [{\citenamefont {Stadler}\ \emph {et~al.}(2018)\citenamefont {Stadler}, \citenamefont {Reddy}, \citenamefont {Basantkumar}, \citenamefont {McGary}, \citenamefont {Estrine}, \citenamefont {Huang}, \citenamefont {Sung}, \citenamefont {Tan}, \citenamefont {Zou}, \citenamefont {Maqableh}, \citenamefont {Shore}, \citenamefont {Gage}, \citenamefont {Um}, \citenamefont {Hein},\ and\ \citenamefont {Sharma}}]{Stadler}%
  \BibitemOpen
  \bibfield  {author} {\bibinfo {author} {\bibfnamefont {B.~J.~H.}\ \bibnamefont {Stadler}}, \bibinfo {author} {\bibfnamefont {M.}~\bibnamefont {Reddy}}, \bibinfo {author} {\bibfnamefont {R.}~\bibnamefont {Basantkumar}}, \bibinfo {author} {\bibfnamefont {P.}~\bibnamefont {McGary}}, \bibinfo {author} {\bibfnamefont {E.}~\bibnamefont {Estrine}}, \bibinfo {author} {\bibfnamefont {X.}~\bibnamefont {Huang}}, \bibinfo {author} {\bibfnamefont {S.~Y.}\ \bibnamefont {Sung}}, \bibinfo {author} {\bibfnamefont {L.}~\bibnamefont {Tan}}, \bibinfo {author} {\bibfnamefont {J.}~\bibnamefont {Zou}}, \bibinfo {author} {\bibfnamefont {M.}~\bibnamefont {Maqableh}}, \bibinfo {author} {\bibfnamefont {D.}~\bibnamefont {Shore}}, \bibinfo {author} {\bibfnamefont {T.}~\bibnamefont {Gage}}, \bibinfo {author} {\bibfnamefont {J.}~\bibnamefont {Um}}, \bibinfo {author} {\bibfnamefont {M.}~\bibnamefont {Hein}},\ and\ \bibinfo {author} {\bibfnamefont {A.}~\bibnamefont {Sharma}},\ }\bibfield  {title} {\bibinfo {title} {Galfenol thin films and
  nanowires},\ }\bibfield  {journal} {\bibinfo  {journal} {Sensors}\ }\textbf {\bibinfo {volume} {18}},\ \href {https://doi.org/10.3390/s18082643} {10.3390/s18082643} (\bibinfo {year} {2018})\BibitemShut {NoStop}%
\bibitem [{Nor(2022)}]{Norcada_2022}%
  \BibitemOpen
  \href {https://www.norcada.com/products/silicon-carbide-membranes/} {} (\bibinfo {year} {2022})\BibitemShut {NoStop}%
\bibitem [{\citenamefont {Huang}(1984)}]{huang1984optical}%
  \BibitemOpen
  \bibfield  {author} {\bibinfo {author} {\bibfnamefont {C.~C.}\ \bibnamefont {Huang}},\ }\bibfield  {title} {\bibinfo {title} {Optical heterodyne profilometer},\ }\href@noop {} {\bibfield  {journal} {\bibinfo  {journal} {Optical Engineering}\ }\textbf {\bibinfo {volume} {23}},\ \bibinfo {pages} {365} (\bibinfo {year} {1984})}\BibitemShut {NoStop}%
\bibitem [{\citenamefont {Servin}\ \emph {et~al.}(2009)\citenamefont {Servin}, \citenamefont {Estrada}, \citenamefont {Quiroga}, \citenamefont {Mosi{\~n}o},\ and\ \citenamefont {Cywiak}}]{servin2009noise}%
  \BibitemOpen
  \bibfield  {author} {\bibinfo {author} {\bibfnamefont {M.}~\bibnamefont {Servin}}, \bibinfo {author} {\bibfnamefont {J.}~\bibnamefont {Estrada}}, \bibinfo {author} {\bibfnamefont {J.}~\bibnamefont {Quiroga}}, \bibinfo {author} {\bibfnamefont {J.}~\bibnamefont {Mosi{\~n}o}},\ and\ \bibinfo {author} {\bibfnamefont {M.}~\bibnamefont {Cywiak}},\ }\bibfield  {title} {\bibinfo {title} {Noise in phase shifting interferometry},\ }\href@noop {} {\bibfield  {journal} {\bibinfo  {journal} {Optics Express}\ }\textbf {\bibinfo {volume} {17}},\ \bibinfo {pages} {8789} (\bibinfo {year} {2009})}\BibitemShut {NoStop}%
\bibitem [{\citenamefont {Klokholm}(1976)}]{klokholm1976measurement}%
  \BibitemOpen
  \bibfield  {author} {\bibinfo {author} {\bibfnamefont {E.}~\bibnamefont {Klokholm}},\ }\bibfield  {title} {\bibinfo {title} {The measurement of magnetostriction in ferromagnetic thin films},\ }\href@noop {} {\bibfield  {journal} {\bibinfo  {journal} {IEEE Transactions on Magnetics}\ }\textbf {\bibinfo {volume} {12}},\ \bibinfo {pages} {819} (\bibinfo {year} {1976})}\BibitemShut {NoStop}%
\bibitem [{\citenamefont {Raghunathan}\ \emph {et~al.}(2009)\citenamefont {Raghunathan}, \citenamefont {Snyder},\ and\ \citenamefont {Jiles}}]{ComparisonMagnetostrictionTechniques}%
  \BibitemOpen
  \bibfield  {author} {\bibinfo {author} {\bibfnamefont {A.}~\bibnamefont {Raghunathan}}, \bibinfo {author} {\bibfnamefont {J.~E.}\ \bibnamefont {Snyder}},\ and\ \bibinfo {author} {\bibfnamefont {D.~C.}\ \bibnamefont {Jiles}},\ }\bibfield  {title} {\bibinfo {title} {Comparison of alternative techniques for characterizing magnetostriction and inverse magnetostriction in magnetic thin films},\ }\href {https://doi.org/10.1109/TMAG.2009.2022327} {\bibfield  {journal} {\bibinfo  {journal} {IEEE Transactions on Magnetics}\ }\textbf {\bibinfo {volume} {45}},\ \bibinfo {pages} {3269} (\bibinfo {year} {2009})}\BibitemShut {NoStop}%
\bibitem [{\citenamefont {Zhao}\ \emph {et~al.}(2002)\citenamefont {Zhao}, \citenamefont {Yalisove}, \citenamefont {Rek},\ and\ \citenamefont {Bilello}}]{zhao2002evolution}%
  \BibitemOpen
  \bibfield  {author} {\bibinfo {author} {\bibfnamefont {Z.}~\bibnamefont {Zhao}}, \bibinfo {author} {\bibfnamefont {S.~M.}\ \bibnamefont {Yalisove}}, \bibinfo {author} {\bibfnamefont {Z.}~\bibnamefont {Rek}},\ and\ \bibinfo {author} {\bibfnamefont {J.~C.}\ \bibnamefont {Bilello}},\ }\bibfield  {title} {\bibinfo {title} {Evolution of anisotropic microstructure and residual stress in sputtered cr films},\ }\href@noop {} {\bibfield  {journal} {\bibinfo  {journal} {Journal of applied physics}\ }\textbf {\bibinfo {volume} {92}},\ \bibinfo {pages} {7183} (\bibinfo {year} {2002})}\BibitemShut {NoStop}%
\bibitem [{\citenamefont {Sebastiani}\ \emph {et~al.}(2020)\citenamefont {Sebastiani}, \citenamefont {Rossi}, \citenamefont {Zeeshan~Mughal}, \citenamefont {Benedetto}, \citenamefont {Jacquet}, \citenamefont {Salvati},\ and\ \citenamefont {Korsunsky}}]{sebastiani2020nano}%
  \BibitemOpen
  \bibfield  {author} {\bibinfo {author} {\bibfnamefont {M.}~\bibnamefont {Sebastiani}}, \bibinfo {author} {\bibfnamefont {E.}~\bibnamefont {Rossi}}, \bibinfo {author} {\bibfnamefont {M.}~\bibnamefont {Zeeshan~Mughal}}, \bibinfo {author} {\bibfnamefont {A.}~\bibnamefont {Benedetto}}, \bibinfo {author} {\bibfnamefont {P.}~\bibnamefont {Jacquet}}, \bibinfo {author} {\bibfnamefont {E.}~\bibnamefont {Salvati}},\ and\ \bibinfo {author} {\bibfnamefont {A.~M.}\ \bibnamefont {Korsunsky}},\ }\bibfield  {title} {\bibinfo {title} {Nano-scale residual stress profiling in thin multilayer films with non-equibiaxial stress state},\ }\href@noop {} {\bibfield  {journal} {\bibinfo  {journal} {Nanomaterials}\ }\textbf {\bibinfo {volume} {10}},\ \bibinfo {pages} {853} (\bibinfo {year} {2020})}\BibitemShut {NoStop}%
\bibitem [{\citenamefont {Basantkumar}\ \emph {et~al.}(2006)\citenamefont {Basantkumar}, \citenamefont {Stadler}, \citenamefont {Robbins},\ and\ \citenamefont {Summers}}]{Basantkumar}%
  \BibitemOpen
  \bibfield  {author} {\bibinfo {author} {\bibfnamefont {R.}~\bibnamefont {Basantkumar}}, \bibinfo {author} {\bibfnamefont {B.}~\bibnamefont {Stadler}}, \bibinfo {author} {\bibfnamefont {W.}~\bibnamefont {Robbins}},\ and\ \bibinfo {author} {\bibfnamefont {E.}~\bibnamefont {Summers}},\ }\bibfield  {title} {\bibinfo {title} {Integration of thin-film galfenol with mems cantilevers for magnetic actuation},\ }\href {https://doi.org/10.1109/TMAG.2006.879666} {\bibfield  {journal} {\bibinfo  {journal} {IEEE Transactions on Magnetics}\ }\textbf {\bibinfo {volume} {42}},\ \bibinfo {pages} {3102} (\bibinfo {year} {2006})}\BibitemShut {NoStop}%
\bibitem [{\citenamefont {Sablik}\ and\ \citenamefont {Jiles}(1993)}]{sablik1993coupled}%
  \BibitemOpen
  \bibfield  {author} {\bibinfo {author} {\bibfnamefont {M.~J.}\ \bibnamefont {Sablik}}\ and\ \bibinfo {author} {\bibfnamefont {D.~C.}\ \bibnamefont {Jiles}},\ }\bibfield  {title} {\bibinfo {title} {Coupled magnetoelastic theory of magnetic and magnetostrictive hysteresis},\ }\href@noop {} {\bibfield  {journal} {\bibinfo  {journal} {IEEE transactions on magnetics}\ }\textbf {\bibinfo {volume} {29}},\ \bibinfo {pages} {2113} (\bibinfo {year} {1993})}\BibitemShut {NoStop}%
\bibitem [{\citenamefont {Shi}\ \emph {et~al.}(2019)\citenamefont {Shi}, \citenamefont {Wu}, \citenamefont {Hu}, \citenamefont {Lu}, \citenamefont {Mu},\ and\ \citenamefont {Zhu}}]{shi2019study}%
  \BibitemOpen
  \bibfield  {author} {\bibinfo {author} {\bibfnamefont {J.}~\bibnamefont {Shi}}, \bibinfo {author} {\bibfnamefont {M.}~\bibnamefont {Wu}}, \bibinfo {author} {\bibfnamefont {W.}~\bibnamefont {Hu}}, \bibinfo {author} {\bibfnamefont {C.}~\bibnamefont {Lu}}, \bibinfo {author} {\bibfnamefont {X.}~\bibnamefont {Mu}},\ and\ \bibinfo {author} {\bibfnamefont {J.}~\bibnamefont {Zhu}},\ }\bibfield  {title} {\bibinfo {title} {A study of high piezomagnetic (fe-ga/fe-ni) multilayers for magnetoelectric device},\ }\href@noop {} {\bibfield  {journal} {\bibinfo  {journal} {Journal of Alloys and Compounds}\ }\textbf {\bibinfo {volume} {806}},\ \bibinfo {pages} {1465} (\bibinfo {year} {2019})}\BibitemShut {NoStop}%
\bibitem [{\citenamefont {Liang}\ \emph {et~al.}(2018)\citenamefont {Liang}, \citenamefont {Dong}, \citenamefont {Celestin}, \citenamefont {Wang}, \citenamefont {Chen}, \citenamefont {Ziemer}, \citenamefont {Page}, \citenamefont {McConney}, \citenamefont {Jones}, \citenamefont {Howe} \emph {et~al.}}]{liang2018soft}%
  \BibitemOpen
  \bibfield  {author} {\bibinfo {author} {\bibfnamefont {X.}~\bibnamefont {Liang}}, \bibinfo {author} {\bibfnamefont {C.}~\bibnamefont {Dong}}, \bibinfo {author} {\bibfnamefont {S.~J.}\ \bibnamefont {Celestin}}, \bibinfo {author} {\bibfnamefont {X.}~\bibnamefont {Wang}}, \bibinfo {author} {\bibfnamefont {H.}~\bibnamefont {Chen}}, \bibinfo {author} {\bibfnamefont {K.~S.}\ \bibnamefont {Ziemer}}, \bibinfo {author} {\bibfnamefont {M.}~\bibnamefont {Page}}, \bibinfo {author} {\bibfnamefont {M.~E.}\ \bibnamefont {McConney}}, \bibinfo {author} {\bibfnamefont {J.~G.}\ \bibnamefont {Jones}}, \bibinfo {author} {\bibfnamefont {B.~M.}\ \bibnamefont {Howe}}, \emph {et~al.},\ }\bibfield  {title} {\bibinfo {title} {Soft magnetism, magnetostriction, and microwave properties of fe-ga-c alloy films},\ }\href@noop {} {\bibfield  {journal} {\bibinfo  {journal} {IEEE Magnetics Letters}\ }\textbf {\bibinfo {volume} {10}},\ \bibinfo {pages} {1} (\bibinfo {year} {2018})}\BibitemShut {NoStop}%
\bibitem [{\citenamefont {Jahjah}\ \emph {et~al.}(2019)\citenamefont {Jahjah}, \citenamefont {Manach}, \citenamefont {Le~Grand}, \citenamefont {Fessant}, \citenamefont {Warot-Fonrose}, \citenamefont {Prinsloo}, \citenamefont {Sheppard}, \citenamefont {Dekadjevi}, \citenamefont {Spenato},\ and\ \citenamefont {Jay}}]{jahjah2019thickness}%
  \BibitemOpen
  \bibfield  {author} {\bibinfo {author} {\bibfnamefont {W.}~\bibnamefont {Jahjah}}, \bibinfo {author} {\bibfnamefont {R.}~\bibnamefont {Manach}}, \bibinfo {author} {\bibfnamefont {Y.}~\bibnamefont {Le~Grand}}, \bibinfo {author} {\bibfnamefont {A.}~\bibnamefont {Fessant}}, \bibinfo {author} {\bibfnamefont {B.}~\bibnamefont {Warot-Fonrose}}, \bibinfo {author} {\bibfnamefont {A.}~\bibnamefont {Prinsloo}}, \bibinfo {author} {\bibfnamefont {C.}~\bibnamefont {Sheppard}}, \bibinfo {author} {\bibfnamefont {D.}~\bibnamefont {Dekadjevi}}, \bibinfo {author} {\bibfnamefont {D.}~\bibnamefont {Spenato}},\ and\ \bibinfo {author} {\bibfnamefont {J.-P.}\ \bibnamefont {Jay}},\ }\bibfield  {title} {\bibinfo {title} {Thickness dependence of magnetization reversal and magnetostriction in fe 81 ga 19 thin films},\ }\href@noop {} {\bibfield  {journal} {\bibinfo  {journal} {Physical Review Applied}\ }\textbf {\bibinfo {volume} {12}},\ \bibinfo {pages} {024020} (\bibinfo {year} {2019})}\BibitemShut {NoStop}%
\bibitem [{\citenamefont {Adolphi}\ \emph {et~al.}(2010)\citenamefont {Adolphi}, \citenamefont {McCord}, \citenamefont {Bertram}, \citenamefont {Oertel}, \citenamefont {Merkel}, \citenamefont {Marschner}, \citenamefont {Schäfer}, \citenamefont {Wenzel},\ and\ \citenamefont {Fischer}}]{AdolphiB}%
  \BibitemOpen
  \bibfield  {author} {\bibinfo {author} {\bibfnamefont {B.}~\bibnamefont {Adolphi}}, \bibinfo {author} {\bibfnamefont {J.}~\bibnamefont {McCord}}, \bibinfo {author} {\bibfnamefont {M.}~\bibnamefont {Bertram}}, \bibinfo {author} {\bibfnamefont {C.-G.}\ \bibnamefont {Oertel}}, \bibinfo {author} {\bibfnamefont {U.}~\bibnamefont {Merkel}}, \bibinfo {author} {\bibfnamefont {U.}~\bibnamefont {Marschner}}, \bibinfo {author} {\bibfnamefont {R.}~\bibnamefont {Schäfer}}, \bibinfo {author} {\bibfnamefont {C.}~\bibnamefont {Wenzel}},\ and\ \bibinfo {author} {\bibfnamefont {W.-J.}\ \bibnamefont {Fischer}},\ }\bibfield  {title} {\bibinfo {title} {Improvement of sputtered galfenol thin films for sensor applications},\ }\href@noop {} {\bibfield  {journal} {\bibinfo  {journal} {Smart materials and structures}\ }\textbf {\bibinfo {volume} {19}},\ \bibinfo {pages} {055013} (\bibinfo {year} {2010})}\BibitemShut {NoStop}%
\bibitem [{\citenamefont {Foner}(1959)}]{foner1959versatile}%
  \BibitemOpen
  \bibfield  {author} {\bibinfo {author} {\bibfnamefont {S.}~\bibnamefont {Foner}},\ }\bibfield  {title} {\bibinfo {title} {Versatile and sensitive vibrating-sample magnetometer},\ }\href@noop {} {\bibfield  {journal} {\bibinfo  {journal} {Review of Scientific Instruments}\ }\textbf {\bibinfo {volume} {30}},\ \bibinfo {pages} {548} (\bibinfo {year} {1959})}\BibitemShut {NoStop}%
\bibitem [{\citenamefont {Bland}\ \emph {et~al.}(1989)\citenamefont {Bland}, \citenamefont {Padgett}, \citenamefont {Butcher},\ and\ \citenamefont {Bett}}]{bland1989intensity}%
  \BibitemOpen
  \bibfield  {author} {\bibinfo {author} {\bibfnamefont {J.}~\bibnamefont {Bland}}, \bibinfo {author} {\bibfnamefont {M.}~\bibnamefont {Padgett}}, \bibinfo {author} {\bibfnamefont {R.}~\bibnamefont {Butcher}},\ and\ \bibinfo {author} {\bibfnamefont {N.}~\bibnamefont {Bett}},\ }\bibfield  {title} {\bibinfo {title} {An intensity-stabilised he-ne laser for measuring small magneto-optic kerr rotations from thin ferromagnetic films},\ }\href@noop {} {\bibfield  {journal} {\bibinfo  {journal} {Journal of Physics E: Scientific Instruments}\ }\textbf {\bibinfo {volume} {22}},\ \bibinfo {pages} {308} (\bibinfo {year} {1989})}\BibitemShut {NoStop}%
\bibitem [{\citenamefont {Takacs}(2001)}]{magnetichysteresismodel}%
  \BibitemOpen
  \bibfield  {author} {\bibinfo {author} {\bibfnamefont {J.}~\bibnamefont {Takacs}},\ }\bibfield  {title} {\bibinfo {title} {A phenomenological mathematical model of hysteresis},\ }\href@noop {} {\bibfield  {journal} {\bibinfo  {journal} {Compel}\ }\textbf {\bibinfo {volume} {20}},\ \bibinfo {pages} {1002} (\bibinfo {year} {2001})}\BibitemShut {NoStop}%
\end{thebibliography}%
\clearpage
\clearpage
\onecolumngrid 
\renewcommand{\thefigure}{S\arabic{figure}}
\renewcommand{\thesection}{S.\arabic{section}}
\renewcommand{\thetable}{S.\arabic{table}}
\renewcommand{\theequation}{S.\arabic{equation}}

\setcounter{figure}{0}
\setcounter{section}{0}
\setcounter{equation}{0}
\setcounter{table}{0}

\begin{centering}
\normalfont\large\textbf{{Supplementary Material for \\
Quantitative Profilometric Measurement of Magnetostriction
in Thin-Films\\}
}%
\end{centering}
\begin{centering}
\vspace{\baselineskip}
Hamish Greenall\authormark{1,2}, Benjamin J. Carey\authormark{1,2}, Douglas Bulla\authormark{4}, James S. Bennett\authormark{3}, Fernando Gotardo\authormark{1,2}, Scott Foster\authormark{4}, and Warwick P. Bowen\authormark{1,2}\\
\vspace{\baselineskip}
\authormark{1}School of Mathematics and Physics, The University of Queensland, St Lucia, Queensland 4067, Australia.\\
\authormark{2}ARC Centre of Excellence for Engineered Quantum Systems, St Lucia, Queensland 4067, Australia.\\
\authormark{3}Centre for Quantum Dynamics, Griffith University, Nathan, Queensland 4072, Australia.\\
\authormark{4}Australian Government Department of Defence Science and Technology, Edinburgh,\\ South Australia 5111, Australia.\\
\end{centering}

\vspace{\baselineskip}

\twocolumngrid 


\section{Paraboloid approximation of a sphere}
\label{supp:paraboloidapprox}
Here we present the paraboloid approximation of a sphere that has been used to approximate the shape of a spherically bowed sample. A centered sphere of radius $R$ in cartesian coordinates is given by the equation
\begin{equation}
	x^2+y^2+z^2=R^2,
\end{equation}
which can be rearranged to
\begin{equation}
	z=\pm R \sqrt{1-\frac{\left(x^2+y^2\right)}{R^2} }.
\end{equation}
In the case that $x^2+y^2 \ll R^2$ the binomial approximation can be used to give 
\begin{equation}
	z=\pm R \left(1-\frac{\left(x^2+y^2\right)}{2R^2}\right).
\end{equation}
Taking the $-$ solution yields
\begin{equation}
	z+R=\frac{1}{2R} \left(x^2+y^2\right)
\end{equation}
which is a paraboloid with center $(0,0, -R)$. 

\section{Difference of two elliptic paraboloids}
\label{sec:differencetwoparaboloids}
The difference of two elliptic paraboloids of the form $z_2=\frac{1}{2}\kappa_{x,2} x^2+\frac{1}{2}\kappa_{y,2} y^2$, $z_1=\frac{1}{2}\kappa_{x,1} x^2+\frac{1}{2}\kappa_{y,1} y^2$ is given by
\begin{equation}
	z=\frac{1}{2}\left(\kappa_{x,2}-\kappa_{x,1}\right) x^2 + \frac{1}{2}\left(\kappa_{y,2}-\kappa_{y,2}\right) y^2,
\end{equation}
which can be written in terms of the change in curvatures $\Delta \kappa_x=\left(\kappa_{x,2}-\kappa_{x,1}\right)$, $\Delta \kappa_y=\left(\kappa_{y,2}-\kappa_{y,2}\right) $ as
\begin{equation}
	z=\frac{1}{2}\Delta \kappa_x x^2 + \frac{1}{2}\Delta \kappa_y y^2 
\end{equation}

\section{deriving the uncertaintyof curvature measurements}
\label{supp:derivingnoiselimit}
In this section we present the derivation of Eq. (\ref{eq:a_variance}), which is an expression for the predicted uncertainty our measurement of the surface curvature $\kappa$. 

\subsection{In 1D}

\begin{figure}[htbp]
	\centering
	\includegraphics[width=0.75\columnwidth]{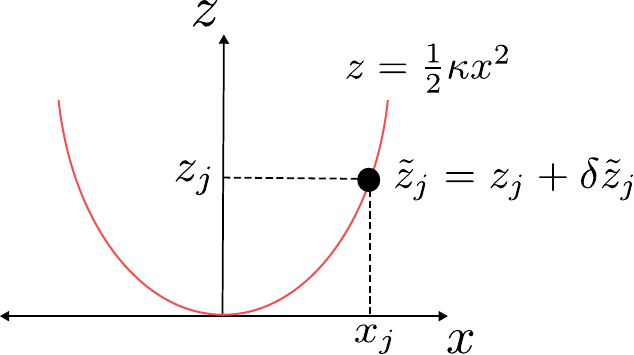}
	\caption{\label{fig:1Dparabola} Diagram showing the 1D model of our measurement of the surface height $z$ to obtain the measurement outcomes $\tilde z_j$.}
\end{figure}

To estimate the uncertainty in the measurement of the surface curvature $\kappa$, we first model the system in 1D. The profilometer measures the height of the parabolic surface $z=\frac{1}{2}\kappa x^2$ at positions $x_j$, to obtain the measurement outcomes $\tilde z_j$, see Fig. (\ref{fig:1Dparabola}). Each measurement has noise $\delta \tilde z_j=\tilde z_j-\langle \tilde z_j\rangle = \tilde z_j-z_j$, where $\langle \tilde z_j\rangle$ denotes the expectation value of $\tilde z_j$. Hence $\tilde z_j$ can be expressed as 
\begin{equation}
	\tilde z_j= z_j +\delta \tilde z_j = \frac{1}{2} \kappa x_j^2 + \delta \tilde z_j.
\end{equation}
Note that for ease of reading, in the main text we have neglected the $\tilde{}$ notation used to denote our estimate of values. To simplify calculations we define $a=\frac{1}{2}\kappa$. For each outcome our measurement of $a$, is
\begin{equation}
	\tilde a_j= \frac{\tilde z_j}{x_j^2}.
\end{equation}
The error in this estimate is
\begin{equation}
	\delta \tilde a_j = \tilde a_j - \langle \tilde a_j \rangle=\frac{\tilde z_j}{x_j^2}-\frac{z_j}{x_j^2}=\frac{\delta z_j}{x_j^2}.
\end{equation}
In which case the signal-to-noise ratio (SNR) of $\tilde a_j$ is 
\begin{equation}
	\mathrm{SNR} (\tilde a_j)=\frac{\langle \tilde a_j^2 \rangle}{(\std(\tilde a_j))^2}=\frac{a^2}{\vari(\tilde a_j)}=\frac{a^2x_j^4}{\vari(\tilde z_j)}.
\end{equation}
Where $\vari(\tilde a_j)=(\std(\tilde a_j))^2$ is the variance. Taking the weighted average of estimates $\tilde a_j$ with an optimal weighting function $w_j$ gives the best estimate of $a$,
\begin{equation}
	\tilde a=\sum_j w_j \tilde a_j,
\end{equation}
where $w_j$ has the property that $\sum\limits_j w_j=1$ . The optimal weighting function is given by
\begin{equation}
	w_j=\frac{\mathrm{SNR}(\tilde a_j)}{\sum\limits_k \mathrm{SNR} (\tilde a_k)} = \frac{a^2x_j^4/\vari(\tilde z_j)}{\sum\limits_k a^2 x_k^4/\vari(\tilde z_k)}=\frac{x_j^4}{\sum\limits_k x_k^4}.
\end{equation}
Where we have assumed the variance $\vari (\tilde z_j)$ is the same for all measurement points $\vari (\tilde z_j)=\vari (\tilde z)$. So
\begin{equation}
	\tilde a = \sum\limits_j \frac{x_j^4}{\sum\limits_k x_k^4} \tilde a_j =\frac{1}{\sum\limits_k x_k^4} \sum\limits_j x_j^2 \tilde z_j.
	\label{eq:tildeasum}
\end{equation}
The variance of this estimate is $\vari(\tilde a) = \langle (\tilde{a}-\langle \tilde a \rangle)^2 \rangle$, which substituting Eq. (\ref{eq:tildeasum}) and simplifying gives 

\begin{equation}
	\vari(\tilde a) = \frac{1}{\left(\sum\limits_k x_k^4\right)^2}\sum_j \sum\limits_k x_j^2 x_k^2 \left\langle \delta \tilde z_j \delta \tilde z_k \right\rangle \label{eq:longsum}.
\end{equation}
In the case of white noise $\left\langle \delta \tilde z_j \delta \tilde z_k \right\rangle = \langle (\delta \tilde z_j)^2 \rangle\delta_{jk}=\vari(\tilde z) \delta_{jk}$  where $\delta_{jk}$ is the Kronecker delta.  Therefore Eq. (\ref{eq:longsum}) evaluates to
\begin{equation}
	\vari (\tilde a) = \frac{\vari(\tilde z)}{\sum_k x_k^4}.
\end{equation}
\subsection{In 2D}
\begin{figure}[htbp]
	\includegraphics[width=0.85\columnwidth]{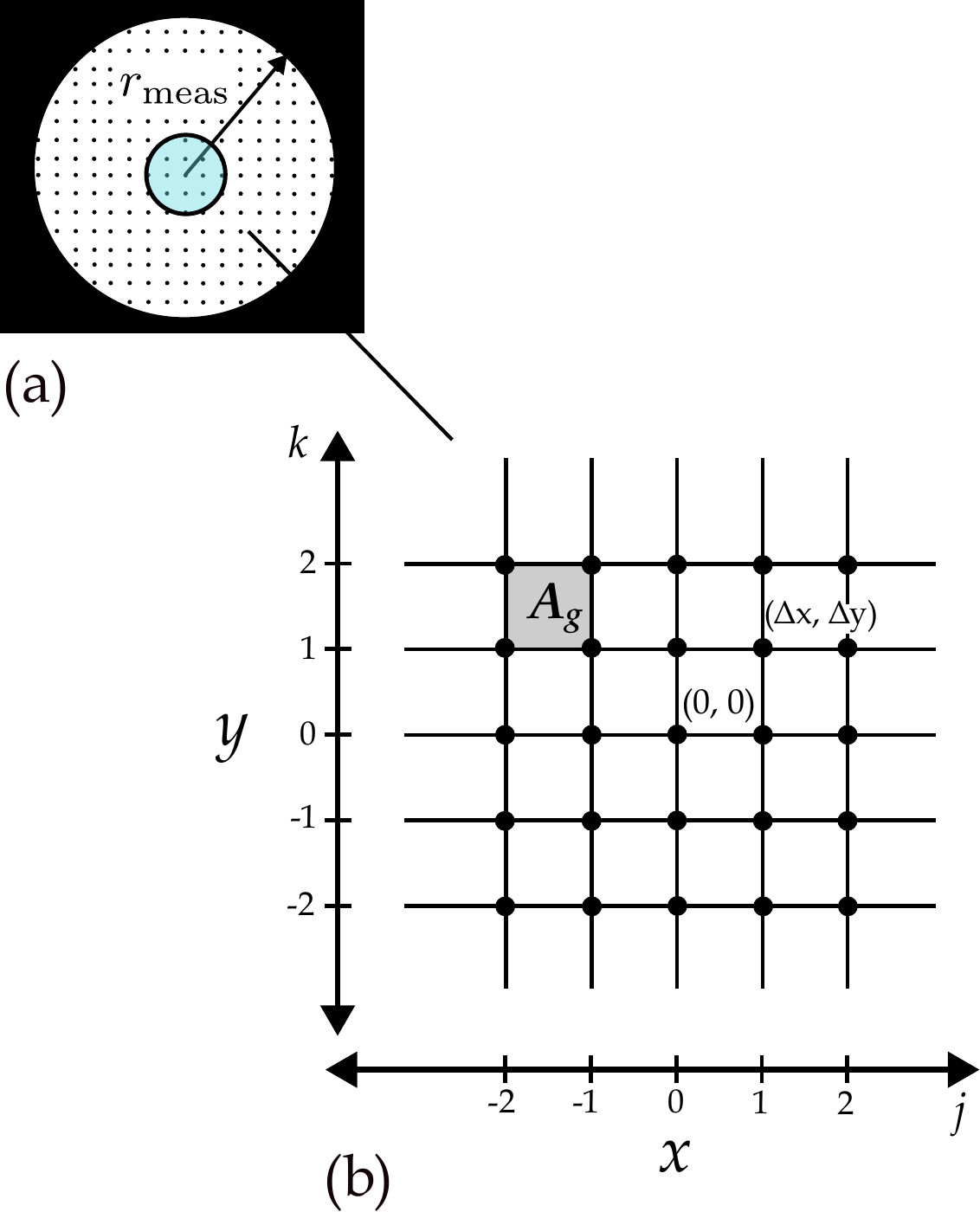}
	\caption{\label{fig:measgrid} Diagram showing the 2D model of our measurement of the surface height. (a) Shows a grid of $N_\mathrm{meas}$ measurements at positions $(x_j, y_j)$ in an area of radius $r_\mathrm{meas}$. (b) Shows a zoomed in section, showing each grid element has size $A_g=\Delta x \times \Delta y$.}
\end{figure}

Moving to 2D, the profilometer measures the height of the surface $z_{j,k}$ in a grid of evenly spaced measurements at positions $r_{j,k}$, across an area of $A_\mathrm{meas}=\pi r_\mathrm{meas}^2$, see Fig. (\ref{fig:measgrid}).
\noindent
Extending $x_j \rightarrow r_{j,k}= \sqrt{x_j^2+y_k^2}$ gives
\begin{equation}
	\vari (\tilde a) = \frac{\vari (\tilde z_j)}{\sum_j \sum_k \left(x_j^2 + y_k^2 \right)^2}.
	\label{eq:asum2d}
\end{equation}
We define $f(x_j, y_k) =(x_j^2+y_k^2)^2$ with $x_j = \Delta x \cdot j$ and $y_k = \Delta y \cdot k$. Each grid element has area $A_g=(\Delta x)^2=(\Delta y)^2 = {\pi r_\mathrm{meas}^2}/{N_\mathrm{meas}}$, where $N_\mathrm{meas}$ is the number of measurement points.  Together this gives
\begin{equation}
	f(j,k)=\left(j^2(\Delta x)^2 +k^2 (\Delta y)^2\right)^2=\left(\frac{\pi r_\mathrm{meas}^2}{N_\mathrm{meas}}\right)^2 \left(j^2+k^2\right).
\end{equation}
The sum in the denominator of Eq. (\ref{eq:asum2d}) can be approximated as the integral
\begin{align}
	\sum_j \sum_k \left(x_j^2 + y_k^2 \right)^2 & \approx \int \int f(j,k)\  \mathrm{d} j \mathrm{d} k \\
	&= \left(\frac{\pi r_\mathrm{meas}^2}{N}\right)^2 \int \int  \left(j^2+k^2\right)^2\ \mathrm{d} j \mathrm{d} k.
\end{align}
Which can be evaluated in polar coordinates with $\ell=\sqrt{j^2 + k^2}$ , $l_{max}= \frac{r_\mathrm{meas}}{\Delta x}=\sqrt{\frac{N_\mathrm{meas}}{\pi}}$ such that
\begin{align}
	\sum_j \sum_k \left(x_j^2 + y_k^2 \right)^2 &\approx \left(\frac{\pi R^2}{N_\mathrm{meas}}\right)^2 \int_0^{2\pi} \int_0^{\ell_{max}} \ell^5 \ \mathrm{d} \ell \mathrm{d} \theta \\
	&= \frac{r_\mathrm{meas}^4 N_\mathrm{meas}}{3}
\end{align}
From which we obtain the $\vari (\tilde a)$ in terms of the $\vari (\tilde z)$ as
\begin{equation}
	\vari (\tilde a) \approx \frac{3 \vari (\tilde z)}{r_\mathrm{meas}^4 N_\mathrm{meas}}.
\end{equation}
Accordingly the uncertainty in our measurement of $\kappa$ is
\begin{equation}
	\std(\tilde \kappa) \approx \sqrt{\frac{6 }{N_\mathrm{meas}}}\frac{\std(\tilde z)}{r_\mathrm{meas}^2}.
\end{equation}

\section{measuring the magnetization hysteresis and investigating the two-layer behaviour}
\label{supp:magnetizationgalf}
Along with magnetostriction measurements we also took magnetization measurements using both a vibrating sample magnetometer (VSM) (Lakeshore 8600 Series) and a Magneto-optical Kerr effect (MOKE) microscope (Evico magnetics). The VSM was used to measure the hysteresis of the films sputtered simultaneously onto silicon chips (0.5~cm $\times$ 0.5~cm) while the MOKE was used to measure the hysteresis of the films on the coverslips.

\subsection{VSM curves and two-layer model}
\label{magnetizationhysteresisloops}
\begin{figure}[htbp]
	\includegraphics[width=0.95\columnwidth]{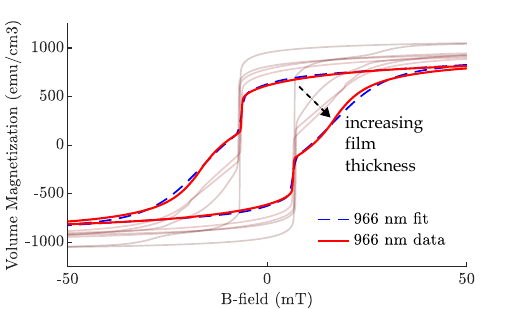}
	\caption{\label{fig:vsm_curves} VSM curves showing the magnetization versus B-field hysteresis loops for the different thicknesses. The two-layer model fit described in Sec. (\ref{subsec:twolayermodel}) shown for the 966~nm film.}
\end{figure}

Fig. (\ref{fig:vsm_curves}) shows the magnetization versus B-field hysteresis curves for galfenol samples of varying thicknesses from 206~nm to 966~nm, measured using VSM. There is a clear qualitative difference between the curves of different thicknesses, whereby, as the film thickness increases the coercivity of the sample increases and a `kink' in the curve becomes more pronounced. As with the magnetostriction measurements this suggests that as the galfenol layer becomes thicker the structure may enter a secondary phase. 

\begin{figure}[htbp]
	\includegraphics[width=0.95\columnwidth]{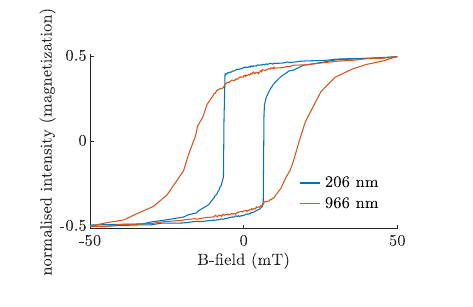}
	\caption{\label{fig:exMOKEloop} Magnetization hysteresis loops measured using MOKE showing the difference in shape and coercivity between the 206~nm and 966~nm films. }
\end{figure}

Further insight into this behavior is provided by the MOKE measurements, which in contrast to the VSM measurements that measure the total magnetization of entire film \cite{foner1959versatile}, measure primarily the first 10 -- 20~nm of the film \cite{bland1989intensity}. The MOKE hysteresis curves show that at the top of the films with total thicknesses larger than 206~nm there is a layer of material with a coercivity of $\sim$16~mT, whereas for the thinner films the coercivity measured is $\sim$6~mT, see Fig. (\ref{fig:exMOKEloop}). These observations reinforce the two-layer model and indicate that the first and secondary phases have coercivities of $\sim$6~mT and $\sim$16~mT respectively.  

In Sec. (\ref{subsec:twolayermodel}) we present a model for this two-layer behaviour by treating the films as composed of two distinct layers, each with a its own thickness, coercivity, remanence, and saturation magnetization. We find that with these coercivities the model approximately fits the VSM curves as shown in Fig. (\ref{fig:vsm_curves}), with a bottom layer thickness of $\sim 175$~nm.

\subsection{\label{subsec:twolayermodel}Two-layer model}

In the two-layer model the total magnetization is taken to be the sum of the magnetizations of each layer scaled by their respective thickness, $t_1$ \& $t_2$. The magnetization hysteresis loop of a single layer, reaching the saturation field, can be described by equations of the form
\begin{align}
	c=\frac{1}{B_C}\mathrm{tanh^{-1}}\left(\frac{M_R}{M_S}\right)\\
	M_+(B)=M_S\mathrm{tanh}\left(c\left(B-B_C\right)\right)\\
	M_-(B)=M_S\mathrm{tanh}\left(c\left(B+B_C\right)\right)
\end{align}where  $M_+$ is the equation for the ascending branch, $M_-$ is the equation for the descending branch, $M_S$ is the saturation magnetization, $c$ is a constant which determines the slope of the hyperbolic tangent, $M_R$ is the remanence and $B_C$ is the coercivity \cite{magnetichysteresismodel}. For the two layer model we take 
\begin{align}
	M_+&=\frac{1}{t_1+t_2} \left(t_1 M_+^{(1)}(B') + t_2 M_+^{(2)}(B'')\right)\\
	M_-&=\frac{1}{t_1+t_2} \left(t_1 M_-^{(1)}(B') + t_2 M_-^{(2)}(B'')\right),
\end{align}
with $^{(1)}$ and $^{(2)}$ denoting the first and second layers, $B'= B\cos(\theta_1)$, $B''= B\cos(\theta_2)$, where $\theta_1$ and $\theta_2$  are the angles of the magnetization axis relative to the applied $B$-field for each layer.

\end{document}